\documentclass[twocolumn,twoside,technote,9pt,final]{IEEEtran}
\usepackage[dvips]{graphicx}
\usepackage{epsfig}
\usepackage{amsmath}
\usepackage{amsopn}
\usepackage{latexsym}
\usepackage{amssymb}
\usepackage{verbatim}
\usepackage{enumerate}
\usepackage{pifont}

\input{prepictex}
\input{postpictex}
\usepackage{color}
\definecolor{turq}{cmyk}{0.25,0,0.5,0}
\definecolor{mycy}{cmyk}{0.2,0,0,0}
\definecolor{mymag}{cmyk}{0,0.3,0,0}
\definecolor{brown}{rgb}{0.8,0.6,0.1}

\newcommand{\sw}{\mbox{\boldmath $s$}}
\newcommand{\chiw}{\mbox{\boldmath $\chi$}}

\newcommand{\be}{\begin{equation}}
\newcommand{\ee}{\end{equation}}
\newcommand{\ba}{\begin{eqnarray}}
\newcommand{\ea}{\end{eqnarray}}

\newcommand{\rf}[1]{\mbox{(\ref{#1})}}

\newcommand{\bi}{\begin{itemize}}
\newcommand{\ei}{\end{itemize}}

\newcommand{\Prob}{\mbox{Pr}}
\newcommand{\lcm}{\mbox{lcm}}
\newcommand{\bd}[1]{\mbox{\boldmath $#1$}}

\newcommand{\df}{\stackrel{\mbox{\scriptsize \rm def}}{=}}
\newcommand{\map}[1]{\stackrel{\mbox{\unboldmath\scriptsize $#1$}}{\mapsto}}

\newcommand{\sta}[3]{\mbox{$ \scriptsize \renewcommand{\arraystretch}{.5}\begin{array}{@{}c@{}c@{}}#1 \\ #2 \\#3\end{array}\renewcommand{\arraystretch}{1}$}}

\newcommand{\mtrx}[4]{\mbox{$\left[\renewcommand{\arraystretch}{0.5}\begin{array}{@{}c@{}c@{}} #1 & \,#2\\ #3 & \,#4  \end{array}\right]\renewcommand{\arraystretch}{1}$}}

\newcommand{\piM}[1]{\mbox{$\left(\bd{#1}^{\rm H} \bd{#1}\right)^{-1}\bd{#1}^{\rm H}$}}

\newcommand{\gm}[1]{\mbox{$\grave{\bd{#1}}$}}
\newcommand{\gv}[1]{\mbox{$\grave{#1}$}}

\newcommand{\colv}[2]{\mbox{$\bd{#1}_{{ \!\{\!#2\!\}}}$}}
\newcommand{\rowv}[2]{\mbox{$_{\{\!#2\!\}}\!\bd{#1}$}}
\newcommand{\diag}{\mbox{diag}}

\newcommand{\titlu}{SISO APP Searches in Lattices with Tanner Graphs}

\begin{document}
\title{\titlu}
\author{Dumitru Mihai Ionescu,~\IEEEmembership{Senior Member,~IEEE},
Haidong~Zhu,~\IEEEmembership{Member,~IEEE}
\thanks{Manuscript received Feb.\ 22, 2006; revised Sep.\ 5, 2011. This work was supported by Nokia
Inc.\ and in part by Olympus Corp.\ of Americas, Inc.\
}
\thanks{Dumitru Mihai Ionescu was with the Nokia Research Center, San Diego, CA USA.
He is now with Olympus Communications Technology of America, Inc., San
Diego, CA 92121 USA (e-mail: dmicmic@ieee.org).}
\thanks{Haidong Zhu was with the Nokia Research Center, San Diego, CA USA, and with Olympus Communications Technology of America, San
Diego, CA 92121. He is now with Qualcomm, Inc., San Diego, CA 92121 USA}

\thanks{Copyright \copyright\ 2011 IEEE. Personal use of this material is permitted. However, permission to use this material for any other purposes must be obtained from the IEEE by sending a request to pubs-permissions@ieee.org.}
} 

\markboth{\titlu---To Appear in IEEE Trans.\ IT}{Ionescu and Zhu}
\maketitle

\begin{abstract}
An efficient, low-complexity, soft-output  detector for general lattices  is presented, based on their Tanner graph (TG) representations. 
Closest-point searches in lattices can be performed as non-binary belief propagation on associated TGs; soft-information output is naturally generated in the process; the algorithm requires no backtrack (cf.\ classic sphere decoding), and extracts extrinsic information.
A lattice's coding gain enables equivalence relations  between lattice points, which can be thereby partitioned in cosets.
Total and extrinsic {\em a posteriori} probabilities at the detector's output further enable the use of soft detection information in iterative schemes. 
The algorithm is illustrated via two scenarios 
 that transmit a 32-point, uncoded  super-orthogonal (SO) constellation for multiple-input multiple-output (MIMO) channels, carved from an 8--dimensional  non-orthogonal lattice $D_4 {\oplus} D_4$:  it achieves maximum likelihood performance in 
 quasistatic fading; and,  performs close to interference-free transmission, and identically to list sphere decoding, in independent fading 
 with coordinate  interleaving and iterative equalization and detection. Latter  scenario outperforms  former despite absence of forward error correction 
  coding---because the inherent lattice coding gain allows for the refining of extrinsic information. The  
  lattice constellation is the same as the one employed in the SO space-time trellis codes first introduced for $2{\times} 2$ MIMO by Ionescu {\em et al.}, 
  then independently by Jafarkhani and Seshadri. 
  Complexity is log-linear  in lattice dimensionality, vs.\ cubic 
  in sphere decoders.
\end{abstract}

\begin{keywords}Belief propagation, closest lattice point search, complexity, iterative decoder, MIMO, soft-output, sphere decoder, Tanner graph 
\end{keywords}
\section{Introduction---Problem Perspective\,and\,Setting\label{sec:intro}}
\PARstart{M}{ultiple} input multiple output (MIMO) transmission has emerged as a strong scenario for future high-speed
wireless communications due to the large capacity potential 
of MIMO channels. Space-time codes that exploit both spatial
diversity and time diversity have been widely proposed as MIMO
modulation in the past decade to achieve reliable transmission.
The importance  of lattice MIMO constellations in constructing  space-time lattice codes was recognized by El-Gamal {\em et al.} \cite{gam:lat} from a diversity-multiplexing tradeoff perspective. Superorthogonal space-time codes---reported first in \cite{IonescuMYL:01}, then in \cite{siw:impcon,siw:import,ses:sup,ion:spa,jaf:sup} (wherein they were dubbed `superorthogonal')---are {\em one} instance of lattice space-time codes; the lattice structure inherent to the superorthogonal constellation was noted by Ionescu and Yan \cite[Sec.\ III]{ion:fad}. 
Its optimality as a linear dispersion constellation was further characterized in \cite{gha:div}, with a generalization beyond $2{\times} 2$ MIMO  proposed in \cite{fer:new}.
Lattice constellations lend themselves to efficient detection algorithms, e.g.\ sphere decoding. Classic sphere decoding \cite{agr:clo}
aims at hard decision, and exhibits a backtrack feature (see footnote~\ref{ftntstpbck}); as summarized below, soft-output variations have been imagined \cite{Hochwald:LSD}, but 
retain the backtrack artifact. Banihashemi and Kschischang employed lattice partitioning \cite{ban:tan} to divide
the infinite lattice  into a finite number of cosets. Each coset
is then labeled by a codeword of a finite Abelian group block
code, called a label code. Tanner graph (TG) representations for
the  label code \cite{ban:tan} opened an opportunity for using  belief propagation (BP)  on lattice labels. 
Sadeghi {\em et al.} \cite{sad:low} construct   `low-density-parity-check  (LDPC) lattices' with large coding gains from nested LDPC codes, then use the lattice TG   to perform a form of message passing;\label{sadeghi}
since `LDPC lattices' already have, by construction, significant coding gain (by virtue of dimensionality), \cite{sad:low} had to solve a pure detection problem---namely, for an uncoded lattice constellation,  albeit one with an inherent (lattice) coding gain---and the message passing simply aimed at finding the closest lattice (constellation) point, without  need or provision for producing soft-output or extrinsic information. \cite{sad:low} exploits a lattice TG from the perspective of an underlying block code, essentially constructing a custom lattice for a given  block code.

The literature on lattice and sphere decoding is very rich---see, e.g., \cite{Hochwald:LSD}--\cite{has:sph}, and the plethora of references therein---with wide interest in the  mathematical formalism of the lattice structure, along with algorithmic and complexity aspects. Advances in lattice theory and sphere decoding were applied to non-MIMO telecommunication problems and reported as early as in the 1990s by Viterbo and Biglieri \cite{vit:uni2}, Viterbo and Boutros \cite{vit:uni1}, Boutros {\em et al.} \cite{bou:goo}. Damen {\em et al.} introduced the sphere decoder to MIMO schemes \cite{dam:lat}; see also \cite{gam:lat,Hochwald:LSD}, along with references mentioned above, and elsewhere herein; in one aspect, Hochwald \cite{Hochwald:LSD} rightfully distinguishes between searching for lattice points that maximize the (detection) likelihood---i.e., solve an integer least squares (LS), or related, problem---and those that maximize the maximum {\em a posteriori} probability (vis-\`a-vis extrinsic information). Soft-output flavors were also pursued by Boutros {\em et al.}\ \cite{bou:sof}, and by Studer {\em et al.}\ \cite{stu:sof}, who compute bit log-likelihood ratios by refining a tree-traversal strategy (cf.\ references in {\em op.\ cit.})---but do not accommodate {\em a priori} information. The potential of lattice and coset codes was recognized in the early work of Forney \cite{for:uni}, who discussed the concept of geometrical uniformity and geometrically uniform constellations. Boutros {\em et al.} discussed an alternate view from the perspective of constellations good in both fading and AWGN channels \cite{bou:goo}; Ionescu and Yan discussed an example of a geometrically uniform  MIMO constellation \cite{yan:geo} with fading resilience (see more in \cite{ion:fad}). Yet another use of lattices as enablers for signal-space (or modulation) diversity (SSD)---in an attempt to convert a fading channel into an AWGN channel---was discussed in \cite{tra:per} in the context of bit interleaved coded modulation (BICM); see also \cite{wan:sys} for some lattice-based space-time block codes. SSD can improve performance in fading channels by boosting diversity order through judicious choice, and use, of the modulator constellation: each group of $N$ consecutive symbols is first mapped to an element of an $N$-dimensional constellation (generally carved from an $N$-dimensional lattice), then a rotation matrix is applied to the lattice constellation in order to maximize the diversity order via the minimum product distance of the lattice \cite{bou:sig}; more on this aspect below. Boutros and Viterbo \cite{bou:sig} showed SSD to render the error performance of an uncoded system insensitive to fading for a sufficiently dimensioned lattice constellation. 
They discussed the idea of interleaving the real coordinates, or components, of points from multidimensional constellations  embedded in some Cartesian product of the complex field $\mathbb{C}$, and pursued an algebraic number-theoretic analysis  to support the conclusion that coordinate interleaving (CI) together with
 constellation rotation can increase diversity---that is, separately from any redundancy  scheme, such as FEC coding. The diversity was quantified  in terms of a coordinate-wise Hamming distance. Viterbo provides---and maintains---some best known constellations for uncoded systems over fading channels \cite{vit:ful,ogg:alg}, obtained from algebraic number theory.
Nevertheless, the SSD problem becomes more complicated when multiple transmit antennas are employed, as coordinates can no longer be observed independently from each other due to the superposition of all transmitted complex symbols at any receive antenna; iterative receivers may have a legitimate role to play here (see e.g.\ \cite{ion:app}). A version of BICM for MIMO channels was examined by Boutros {\em et al.} \cite{bou:bit}. Ionescu {\em et al.} quantified the effect of CI on mutual information in MIMO communications \cite{ion:int}. 
In fading MIMO channels the merit of the product distance---related to coordinate-wise Hamming distance---was posed by Tarokh as a design criterion \cite{tar:spa}; Ionescu \cite{ion:spa} outlined an inherent interdependence between product and Euclidean distances. Studer {\em et al.}\ \cite{stu:sof} show that VLSI implementation of a MIMO soft-output sphere decoder, with channel matrix regularization, is possible at only 58\%  area penalty vs.\ a hard-decision sphere decoder---which in \cite{stu:sof} is a Schnorr-Euchner version \cite{agr:clo} of Pohst's algorithm \cite{poh:com} (finds the correct layer {\em earlier} in the search).
   
Notably, detecting and/or decoding lattice constellations play(s) a key role in the aforementioned problems---SSD included. The problem of efficiently searching through a lattice becomes central in making good use of the signal space. Since the search target is a finite, discrete set of candidate points, unconstrained LS methods, e.g.\ matrix pseudoinverse, cannot solve the problem---albeit, they can guide it, provided that the lattice structure, or its geometrical shape, is not destroyed while deriving a sufficient search statistic; examples are zero-forcing (inherently noise-enhancing), nulling and cancellation (or decision feedback), nulling and ordered cancellation (VBLAST \cite{fos:lay,has:eff}). Optimal lattice detection, or decoding, is a {\em constrained} search problem, aiming to solve an {\em integer least-squares} type of problem, and it is crucial to avoid an exponentially  complex (in lattice dimensionality) exhaustive search. Sphere decoding strives to achieve a polynomial complexity---at least when averaged over noise and lattice generators \cite{has:sph}---by reducing the search space; a widely used philosophy is to enact, and manage, a searching radius during an iterative search process that progresses from a one-dimensional subspace to increasingly higher dimensional subspaces, until the algorithm reaches and ranks one or more lattice point(s) \cite{vit:uni2,Hochwald:LSD,has:sph}; see Agrell {\em et al.} \cite{agr:clo} for an informative survey of classic sphere decoding algorithms, including the one due to Pohst \cite{poh:com}, Fincke and Pohst \cite{fin:imp}. The list sphere decoding (LSD) algorithm  adopts a slight variation on traditional sphere decoding,  in that the radius is purposely prevented from decreasing during search---in light of the possibility that the closest lattice point may not be the one that maximizes {\em a posteriori} extrinsic information \cite[Sec.\ III.B]{Hochwald:LSD}, which is the real search target. In its essence, sphere decoding is more concerned with discovering lattice points within a search sphere (parallelogram in Kannan's algorithm \cite{kan:imp}) than with choosing, and/or managing, a search radius. Whenever such computations {\em cannot} be completed beforehand---e.g., when the channel matrix is inevitably lumped with the lattice generator \cite{Hochwald:LSD}, and whereby the overall lattice geometry changes with the channel use---finding the lattice covering radius is NP-hard, while the so-called Babai estimate \cite{gro:geo} may allow inefficiently many points. \cite{has:sph} advocates that the sphere radius be chosen based on noise variance alone, something also suggested in  \cite{vit:uni2,vit:uni1}; however, \cite[Sec.\ IV.A]{has:sph} does allow for a provision to adjust that radius (should no lattice point be found) by increasing a confidence interval. This suggests a correlation between radius and lattice geometry---albeit one discoverable iteratively (by adjusting the confidence interval).

The state-of-the-art of sphere decoding will be placed in some additional perspective in Sec.~\ref{sota}, as part of a comparison between classic approaches vs.\ the method proposed herein, and vis-\`{a}-vis complexity---which turns out to be data-dependent (see \cite[Sec.\ III-B]{has:sph} for a self-contained argument).
Hassibi and Vikalo argue \cite[Sec.\ III-B.1]{has:sph} that the mean algorithmic complexity of Fincke and Pohst's sphere decoding is exponential in lattice dimensionality---in the sense of  mean number of visited points, and given an arbitrary,  fixed, lattice generating matrix, with arbitrary noise realization; \cite{has:sph} includes a closed-form expression for an {\em expected} complexity measure---averaged over noise and lattice generating matrices---arguing that the expected complexity of classic sphere decoding is polynomial (cubic) in lattice dimensionality. This is consistent with other simulation-driven observations (when inherently averaging over lattices \cite{Hochwald:LSD}).

In light of the above, the sequel takes a novel, qualitatively different approach to soft-output (closest) point search in lattices, via a form of  BP on a lattice.
Orthogonality, or near-orthogonality, of the underlying lattice is not an enabler or facilitator for the algorithm, which accepts a general lattice, and can identify the necessary structure for partitioning and labeling---either in real-time or beforehand; see \ref{TG}. Nonetheless, certain features of lattices that are deemed useful in practice, e.g.\ cycle-free TGs \cite{ban:tan}, remain desirable. 

In order to establish that the algorithm proposed in Sec.\ \ref{sec:detector} is well-defined with respect to (w.r.t.) complexity---yet without resorting to some form of expectation (as was pursued in \cite[Sec.\ III-B]{has:sph})---we first illustrate the existence of lattices for which complexity is no worse than  $O(m\log m)$, i.e.\ non-exponential in the lattice dimensionality $m$, then bound complexity for arbitrary lattices. 

The idea is placed in perspective vis-\`a-vis known approaches in Sec.\ \ref{sota}, which includes the discussion on complexity.
The coding gain inherently associated with a lattice enables deeper structural relations  between subsets of lattice points, which can be thereby associated via an equivalence relation for detection purposes---while helping to constrain complexity in the process; a Markov model is constructed for the lattice, completing a structured framework  for  (i) processing  label probabilities supplied by message passing on the TG, and (ii) generating both total and extrinsic  {\em a
posteriori} probability (APP) at the detector's output; more in  Sec.\ \ref{sec:adv}. Backtracking\footnote{\protect\label{ftntstpbck}\protect\hspace{0pt}Backtracking refers to a known artifact of the sphere-decoding algorithm, whereby the progression from the initial to  final coordinates that are being searched within a sphere must temporarily revisit a previous coordinate  \cite{agr:clo}, \cite[Sec.\,III]{Hochwald:LSD}---either the immediately previous one, or earlier ones if needed---because the search boundaries set by the sphere radius are being exceeded for a tentative point, or no valid lattice point is found within their limits.} is not needed.
In non-AWGN channels, for each channel use, a  
minimum mean square error (MMSE) interference cancellation (IC) filter bank
may be employed to remove the channel effects on the lattice generator matrix \cite{xiaodong:mmse}---just so as to isolate the lattice generator from channel fading, which can in turn alter (sometimes irreversibly) the geometrical shape of the lattice. 
This is not a limitation in principle \label{nolimitatinprinc} of the proposed search philosophy, since the discovery (decomposition) of the lattice structure can be processed in real time (if preferred, or not otherwise deemed  more expensive practically); the search on the lattice TG can then be carried out as proposed---at the cost of additional complexity at the receiver, in response to the fact that the underlying lattice changes with every use of the channel. 
Real-time decomposition of the lattice was advocated and practiced in \cite{Hochwald:LSD,vit:uni1,stu:sof}. Note also the discussion in \cite{ion:fad} concerning MIMO lattice constellations that are actually robust, resilient to fading (their geometrical shape is recoverable even after fading and CI); we conjecture that such constellations might be better understood, and proven to possess desirable traits.

While Section~\ref{nonbinarybp} describes a sum-product algorithm, low-complexity versions (e.g.\ min-sum, see \cite{sad:low}) are possible with the known benefits. 
APP computation
in a soft-input soft-output  (SISO) module that exchanges information with  BP 
enables iterative detection and decoding. 
The 
algorithm 
is illustrated on detecting
a superorthogonal, geometrically uniform, space-time lattice  \cite{ion:fad}---in both
quasistatic fading, and  a coordinate interleaved \cite{ion:int} scenario.
Lowercase, respectively uppercase bold letters denote vectors and matrices; $a_i$ denotes
the $i$-th element of  $\bd{a}$; $\bd{a}_i$,  $a_{ij}$ denote the $i$-th column, respectively $ij$-th
element of  $\bd{A}$; $\rm H$, $\rm T$, $\left\langle \cdot,\cdot\right\rangle $ denote (complex conjugated) transposition and inner product, $\log$ denotes base-2 logarithm, unless otherwise specified.

\subsection{Lattices, Detection Models, MIMO Channels\label{sec:lattdetmodel}}
Consider MIMO wireless transmission with $N_t$, $N_r$ transmit, respectively
 receive antennas  in Rayleigh flat fading. 
\subsubsection{Complex, MIMO, Rayleigh Flat Fading Channels }
Assume the channel to be constant over a block of $T$ MIMO
channel uses and changes independently across blocks; then
\be \textstyle \label{e:MIMO}
\bd{Y} = \sqrt{{1}/{N_t}}\bd{S}\bd{\bar{H}}+ \bd{N} \ee where
$\bd{Y} \in \mathbb{C}^{T \times N_r}, \bd{\bar{H}} \in
\mathbb{C}^{N_t \times N_r}, \bd{S} \in \mathcal{A}^{T \times
N_t}$, $\bd{N} \in \mathbb{C}^{T \times N_r}$ are {\em arrays}
of received signals, channel gain coefficients, transmitted
signals and additive noises, respectively. The elements of
$\bd{N}$ are i.i.d. zero-mean complex-valued Gaussian random
variables with variance $N_0/2$ per dimension, i.e., $n_{ij} \sim
\mathcal{CN}(0, N_0)$. The
channel  coefficients between the $i$-th transmit  and
the $j$-th receive antennas are $\bar{h}_{ij} \sim \mathcal{CN}(0,1)$,  pairwise independent; $\bd{S}$ models the transmitted symbols chosen from
alphabet $\mathcal{A}$\footnote{Different alphabets could be used
on distinct transmit antennas, e.g.\ $\mathcal{A}_j$ on the  $j$-th transmit antenna; alphabets
$\mathcal{A}_j$ could differ, e.g., when identical
constellations are assigned with unequal powers to 
transmit antennas---such could be accommodated
but secondary in importance to purpose of this work.};
$s_{ij}\in \mathbb{C}$ is radiated  from the $j$-th transmit
antenna during the $i$-th channel use. By enforcing the power
constraint 
\be\textstyle{ {\rm E} \left\{|\!|\!|\bd{S}|\!|\!|^2/T
\right\} \leq N_t, }\ee 
where $|\!|\!|\!\cdot\!|\!|\!|$ and $\rm E\{\!\cdot\!\}$ denote 
Euclidean matrix norm and   expectation,
the {\em average} signal-to-noise (SNR) ratio per receive antenna
is $1/N_0$.
\rf{e:MIMO} 
accommodates various setups, which include the case $T\!\!=\!\!1$ that allows for independent (rather than block) fading.
Arrays $\bd{S}$ may have a certain structure, e.g.\
 representing space-time codematrices; or, they  may simply be 
arrays of unrelated values obtained after interleaving  real
coordinates of structured matrices (Sec.~\ref{itercoord}) then 
forming new complex valued arrays by pairing up scrambled
coordinates.

\subsubsection{Equivalent Real-Valued Transmission Model}
It is  convenient to
introduce equivalent real-valued  models 
via the  isomorphisms $\mathcal{I}: \mathbb{C}^{M} {\longmapsto} \mathbb{R}^{2M {\times} 1}$,  $\phi: \mathbb{C}^{M {\times} N} {\longmapsto} \mathbb{R}^{2MN {\times} 1}$ 
\ba \mathcal{I}(\bd{a}) &
\df & [\Re(\bd{a})^{\rm T} \, \Im(\bd{a})^{\rm T}]^{\rm T}, \label{vecisom} \\
\phi(\bd{A})& \df & [\mathcal{I}(\bd{a}_1)^{\rm T} \cdots \mathcal{I}(\bd{a}_{N})^{\rm T}]^{\rm T}, \label{matisom} \ea 
where $\bd{a} \in \mathbb{C}^{M \times 1} $
and $\bd{A} = [\bd{a}_1\, \ldots \, \bd{a}_N] \in \mathbb{C}^{M \times
N}$. The real-valued transmission model that is equivalent to
(\ref{e:MIMO}) is  
\be \label{e:Gamal} \textstyle \bd{y}^{\rm c} = \bd{H}^{\rm c}
\bd{x} + \bd{n}^{\rm c} \ee 
where $\bd{y}^{\rm c} \df \phi(\bd{Y}^{\rm T})$, $\bd{n}^{\rm c}\df \phi(\bd{N}^{\rm T})$, $\bd{x}\df\phi(\bd{S}^{\rm T})$ and $ \bd{H}^{\rm c}\df
\bd{I}_T \otimes \left( \mtrx{ \Re(\bd{\bar{H}}^{\rm T})}
{-\Im(\bd{\bar{H}}^{\rm T})}{\Im(\bd{\bar{H}}^{\rm T})}{\Re(\bd{\bar{H}}^{\rm T})}
\right)$. Note that $\bd{H}^{\rm c}$ is a $2N_rT \times 2N_tT$ block-diagonal real channel
matrix consisting of $T$ identical diagonal replicas the same $2N_r \times 2N_t$ matrix ($\bd{I}_T$ is the identity matrix of
dimension $T$ and $\otimes$ denotes the Kronecker product).  A
similar model has been reported in \cite{gam:lat}.
Define, further,  $\bd{y} = \phi(\bd{Y})$; 
  via \rf{matisom}, $\bd{y}$ is some permutation $\pi$ of $\bd{y}^{\rm c}$, since
$\bd{y}$,  $\bd{y}^{\rm c}$  are $\phi$-isomorphisms of $\bd{Y}$ and 
$\bd{Y}^{\rm T}$. If $ {\pi}(\bd{H}^{\rm c})\df \bd{H}$ denotes a row permutation of $\bd{H}^{\rm c}$ by ${\pi}$ then  $\bd{y}$ relates to $\bd{y}^{\rm c}$ as:
\be \bd{y} \! = \! {\pi}(\bd{y}^{\rm c})    \!            =  \! {\pi}(\bd{H}^{\rm c} \bd{x} + \bd{n}^{\rm c}) \! = \! {\pi}(\bd{H}^{\rm c}) \bd{x} + {\pi}(\bd{n}^{\rm c}) \! = \! \bd{H} \bd{x} + \bd{n}, \label{e:realtxmodel} \ee
Models (\ref{e:realtxmodel}), (\ref{e:Gamal}) are interchangeable, both
equivalent to the MIMO model in \rf{e:MIMO}; 
 (\ref{e:realtxmodel}) will be preferred below as it aligns with 
 \cite{ion:fad}---in order to relay certain  properties of the super-orthogonal
set   used to illustrate the algorithm of Section \ref{sec:detector}. 

\subsubsection{Space-Time Lattice Codes}\label{sec:stlattcd}
An $m$-dimensional real lattice $\Lambda$ is a discrete additive
subgroup of $\mathbb{R}^m$ defined as $\Lambda = \{ \bd{B}\bd{u} :
\bd{u} \in \mathbb{Z}^{m} \}$ where the real matrix $\bd{B}$ of
size $m \times m$  is the generator matrix of $\Lambda$
\cite{gam:lat}. A lattice code $\mathcal{C}(\Lambda, \bd{u}_0,
\mathcal{R})$ is the finite subset of the lattice translate
$\Lambda + \bd{u}_0$ inside some shaping region $\mathcal{R}$,
i.e., $\mathcal{C}(\Lambda, \bd{u}_0, \mathcal{R}) = \{\Lambda +
\bd{u}_0 \} \cap \mathcal{R}$, where $\mathcal{R}$ is a bounded
region of $\mathbb{R}^m$ \cite{gam:lat}, and $\bd{u}_0$ need not be in $\Lambda$.  A space-time coding
scheme with a space-time code matrix set $\mathcal{S}$, such that $\phi(\bd{S}^{\rm T}) \in \mathbb{R}^m$ \ for all $\bd{S} \in \mathcal{S}$, is a lattice space-time code if the
$m$-dimensional image of $\mathcal{S}$ via the isomorphism $\phi$ is a lattice code
$\mathcal{C}(\Lambda, \bd{u}_0, \mathcal{R}),$ i.e.,
$\phi(\{\mathcal{S}^{\rm T}\}) = \left\{ \{\bd{B}\bd{u}: \bd{u} \in \mathbb{R}^m
\} + \bd{u}_0 \right \}\cap \mathcal{R}$. Many well-known
space-time modulation schemes in the literature indeed can be
treated as space-time lattice codes. Two relevant examples of
space-time lattice codes are given below and in Appendix \ref{appndx}.
\newtheorem{ex2}{Example}
\begin{ex2}
\label{ex2}
[{\em Super-orthogonal space-time lattice codes are carved from $D_4 {\oplus} D_4$}] 
A super-orthogonal space-time code is constructed \cite{ion:fad} by expanding  a (generalized) orthogonal design \cite{tir:squ}, which in turn is obtained as a linear combination of matrices similar to the linear dispersion codes \rf{complexlincomb}, \rf{reallincomb}, with expansion coefficients derived from a complex vector \sw;  the difference from a linear dispersion code is that the latter matrices verify  an additional constraint (see \cite[eqs.\ (2), (3)]{ion:fad}). A 32-element super-orthogonal codebook  for QPSK and $T{=}N_t {=} 2$ was described in 
\cite{IonescuMYL:01,ion:spa} and later, independently, in \cite{siw:impcon,siw:import,ses:sup,jaf:sup}. 
A generic codematrix  $\bd{S}$
is\footnote{\label{noteA} Definition \rf{vecisom} of the isomorphism $\mathcal I$ from a complex vector to a real vector differs slightly from  \protect\cite{ion:fad}, where it was defined by interlacing the real and imaginary parts; i.e., in \protect\cite{ion:fad}, if $\sw = [z_1, \ldots ,
z_K]^{\rm T} \in {\mathbb C}^{K}$ then  ${\cal I}(\sw)= \chiw \df [\Re \{z_1\}, \Im \{z_1\}, \ldots, \Re \{z_K\}, \Im \{z_K\}]^{\rm T}$---rather than keeping the real (and imaginary) parts together as  done in eq.\ \rf{vecisom}. This is the reason for swapping the second and third matrices in eqs.\ \rf{C}, \rf{Cprime} relative to \protect\cite[Sec.\ III]{ion:fad}.} \cite{ion:fad}
\be \textstyle \bd{S} = \sum^{3}_{l=0} \chi_l \bd{C}_l +
\sum^{3}_{l=0} {\chi_l}' \bd{C}'_l,\ee 
\be \chi_l \not= 0 \Rightarrow \chi'_l = 0 {\ \rm and \ } \chi'_l \not= 0 \Rightarrow \chi_l = 0, \forall l\ee
Above, $\chi_l$ and
${\chi'_l} (l=0,1,2,3)$ are either $1,-1$, or 0 and the nonzero values are real parts of complex elements from a complex QPSK constellation; the two sets of real coefficients $\chi_l$ and
${\chi'_l} (l=0,1,2,3)$ are not simultaneously nonzero, i.e.\ either
all $\chi_l$s or all ${\chi'_l}$s vanish. Per \cite{ion:fad}, the super-orthogonal matrix codebook is embedded into an 8-dimensional real vector space obtained as a direct sum of two 4-dimensional real vector spaces\footnote{In the superorthogonal construction   the two 4-dimensional components of the direct sum are reflection symmetries (around  origin) of one another \cite{yan:geo}.}. The  matrix sets $\bd{C}_l$, $\bd{C}'_l$ are basis matrices in the  component vector spaces that form the direct sum:
\be \bd{C} =\left\{
\mtrx{1}{0}{0}{-1}, \mtrx{0}{1}{1}{0}, \mtrx{i}{0}{0}{i},  \mtrx{0}{-i}{i}{0}\right\}
\label{C},\ee 
 \be \bd{C}' = 
\left\{\mtrx{1}{0}{0}{1}, \mtrx{0}{-1}{1}{0},  \mtrx{i}{0}{0}{-i}, \mtrx{0}{i}{i}{0}\right\}. \label{Cprime}\ee
$\bd{S}$ is isomorphic with $\bd{x}=\phi(\bd{S}^{\rm T})$, which verifies
\be \label{e:lattice_super}\bd{x} =
\phi(\bd{S}^{\rm T}) = \bd{\Gamma} \bd{\chi}_{\oplus} \ee 
where
$\bd{\chi}_{\oplus} =
[\chi_0,\cdots,\chi_3,{\chi_0}',{\chi_1}',{\chi_3}']^{\rm T} = [\bd{\chi}^{\rm T} \bd{\chi}'^{\rm T}]^{\rm T} \in {\mathbb R}^8$ is a direct sum of two 4-dimensional vectors,
 $\bd{\Gamma} = [\bd{\Gamma}_1\, \bd{\Gamma}_2]$ is a $8 \times
8$ real matrix and $\bd{\Gamma}_1 =
\left[\phi(\bd{C}_0^{\rm T}),\cdots,\phi(\bd{C}_3^{\rm T})\right]$ and
$\bd{\Gamma}_2 =
\left[\phi(\bd{C}_0^{'\rm T}),\cdots,\phi(\bd{C}_3^{'\rm T})\right]$,
respectively. It also follows from \cite{ion:fad} that $\bd{\Gamma}$ is proportional with a unitary matrix via $\bd{\Gamma}\bd{\Gamma}^{\rm H}=2\bd{I}_8$. 
As \sw\ takes values from a QPSK constellation $\{\pm 1 \pm j\}$, $j= \sqrt{-1}$, the nonzero realizations of either of the  vectors $\bd{\chi}$, $\bd{\chi}'$ are the sixteen 4-dimensional real vectors with elements $\pm 1$; i.e., either $\bd{\chi}_{\oplus} = [\bd{\chi}^{\rm T} [0 \, 0 \, 0 \, 0 \,]^{\rm T}]^{\rm T}$ or $\bd{\chi}_{\oplus} = [[0 \, 0 \, 0 \, 0 \,]^{\rm T} \bd{\chi}'^{\rm T}]^{\rm T}$.

Since $\bd{\chi}_{\oplus} \in \mathbb{Z}^8$, the
vector $\bd{x}$ is seen  to be from some lattice $\Lambda$ with generator matrix $\bd{\Gamma}$, via \rf{e:lattice_super}. 
One recognizes that $\bd{\chi}_{\oplus}$ belongs to a direct sum a two 4-dimensional checkerboard lattices. Indeed, consider the lattice $L\df D_4 \oplus D_4$; i.e., a point $\left[ \lambda_1 \, \lambda_2 \ldots \lambda_8 \right]$ in $D_4 \oplus D_4$ has the property that $\left[ \lambda_1 \, \lambda_2 \, \lambda_3 \, \lambda_4 \right]$, $\left[ \lambda_5 \, \lambda_6 \, \lambda_7 \, \lambda_8 \right]$ are from $D_4$. Let $\left[ d_1 \, d_2 \, d_3 \, d_4 \right]$ denote a point in the second shell of $D_4$, i.e.\ satisfying $\sum_{i=1}^4 d_i^2=4$. There are twenty four points in the second shell of $D_4$, of which exactly sixteen will satisfy $|d_i|=1$; denote this set by $\mathcal D$.  If $\bd{B}$ is the $4\times 4$ generator matrix of $D_4$ then $D_4 \oplus D_4$ has generator matrix \mtrx{\bd{B}} {\bd{0}_{4\times 4}} {\bd{0}_{4\times 4}}{\bd{B}}. Then $L =L_1 \oplus L_2$, where $L_1$ and  $L_2$ have generator matrices $\left[ \bd{B} \, \bd{0}_{4\times 4} \right]$ and respectively $\left[ \bd{0}_{4\times 4} \, \bd{B}  \right]$. Both $L_1$ and $L_2$ are isomorphic with $D_4$. $L_1$ contains the sixteen points in the set  $\left\{ \left[ \bd{c}^{\rm T} \, [0 \, 0 \, 0 \, 0]^{\rm T} \right]^{\rm T} | \bd{c} \in {\mathcal D} \right\}$, and $L_2$ contains the sixteen points in the set  $\left\{ \left[ 0 \, 0 \, 0 \, 0]^{\rm T} \,  \bd{c}^{\rm T} \right]^{\rm T} | \bd{c} \in {\mathcal D} \right\}$. Note that the nonzero realizations of 
$\bd{\chi}$, $\bd{\chi}'$ are the sixteen points in the second shell of $D_4$ having unit magnitude real coordinates; thereby, $\Lambda =\Lambda_1 \oplus \Lambda_2$ where $\Lambda_i$ is isomorphic with $L_i$, $i=1,2$, and $\bd{\chi}_{\oplus}$ is from a direct sum of two 4-dimensional checkerboard lattices. See \rf{e:D4} for a generator matrix for 
$D_4$.

It follows from \rf{e:lattice_super} that $\bd{x} =\phi(\bd{S}^{\rm T})$ can be written as 
\be \label{e:lattice_superbis}\bd{x} = \bd{\Gamma} \bd{\chi}_{\oplus} = \bd{\Gamma} \mtrx{\bd{B}} {\bd{0}_{4\times 4}} {\bd{0}_{4\times 4}}{\bd{B}} \bd{u}, \  \bd{u} = \left[ u_1 \, \ldots \, u_4\right]^{\rm T} \!\! \in \mathbb{Z}^{4}\ee
where $\bd{B}$ is the generator matrix of the checkerboard lattice $D_4$, given in \rf{e:D4}. Thereby, $\bd{x}$ can be viewed as being from a lattice with generator matrix $\bd{\Gamma} \mtrx{\bd{B}} {\bd{0}_{4\times 4}} {\bd{0}_{4\times 4}}{\bd{B}} = \left[ \bd{\Gamma}_1 \bd{B} \ \bd{\Gamma}_2 \bd{B}\right]$.
Now the real equivalent transmission model in eq.\  \rf{e:realtxmodel} becomes 
\be
\label{e:superothogonal}\bd{y}\! = \!\bd{H}\bd{x} + \bd{n}\! =\!
\bd{H}\bd{\Gamma}\bd{\chi}_{\oplus} + \bd{n}\! =\!
\bd{H}_{\oplus}\bd{\chi}_{\oplus} + \bd{n} \!=\! \bd{H}_{\oplus}\bd{B u}+ \bd{n}\ee 
where the second
equality is obtained  according to (\ref{e:lattice_super}), and
$\bd{H}_{\oplus} \df  \bd{H}\bd{\Gamma}$. Note that in
\cite{ion:fad} the transmission model for the same
super-orthogonal space-time code is (see footnote \protect\ref{noteA}): \be \bd{y}_{\oplus} =
\bd{G}_{\oplus} \bd{\chi}_{\oplus} + \bd{n}_{\oplus}. \ee 
It can be verified that 
$\bd{G}_{\oplus} =  \mtrx{\bd{H}\bd{\Gamma}_1} {\bd{0}_{4\times 4}} {\bd{0}_{4\times 4}}{\bd{H}\bd{\Gamma}_2}$. 
Furthermore,
the matrix $\bd{G}_{\oplus}$ was shown in \cite{ion:fad} to be proportional with
a unitary matrix, i.e., $\bd{G}_{\oplus}
{\bd{G}_{\oplus}}^{\rm H} = \alpha \bd{I}$. Denote
$\bd{H}^1_{\oplus}\df \bd{H}\bd{\Gamma}_1$ and $\bd{H}^2_{\oplus}\df \bd{H}\bd{\Gamma}_2$.
Then, $\bd{H}^k_{\oplus}$, $k=1,2$, is unitary
up to a scalar $\alpha\in \mathbb{R}$, i.e.\
\be {\bd{H}^k_{\oplus}}^{\rm H}
\bd{H}^k_{\oplus} = \alpha \bd{I},~ k=1,2. \label{Hoplusunitary}\ee
\end{ex2}

\section{Soft-output reduced search in lattice TG} \label{sec:detector}
While lumping a channel matrix with some (equivalent) generator matrix---as  in \rf{e:superothogonal}---might be tempting,  the new generator matrices $\bd{H}\bd{\Gamma}$, $\bd{H}\bd{\Gamma}\bd{B}$ may yield  large label coordinate alphabets  (see \cite[Sec.\ \protect\ref{TG}]{ban:tan}) for  random $\bd{H}$---unless some form of basis reduction can be devised. The concept is more easily  illustrated by  removing the effect of the channel matrix $\bd{H}$ via an equalization step, then dealing with the underlying lattice separately. This is the approach taken below.

A novel soft-information detection algorithm for lattice
space-time constellations is introduced below. Detection is performed in
two  stages: linear minimum mean square error (LMMSE) filtering, and BP
on a lattice. In the former, a finite impulse response (FIR) LMMSE filter bank is used
to remove the  effect of the channel; this step is equivalent to Wang and Poor's parallel IC \cite{xiaodong:mmse}, refined by T\"uchler {\em et al.} \cite{tuc:tur}. Lattice redundancy is subsequently exploited by a novel lattice
detector based on a TG representation. 

\subsection{Interference Canceling MMSE (IC-MMSE) Soft Equalizer} \label{sec:equalizer}
The equivalent real transmission model in \rf{e:realtxmodel} applies.
The goal of the MMSE soft equalizer (filter bank in Fig.\ \ref{icmmse}) is to remove the effect of the
channel $\bd{H}$, and provide a soft estimate of transmitted lattice
$\bd{x}$ so as to minimize the cross-antenna interference due to
other coordinates $\left\{x_l\right\}_{l=1,l\not=i }^{2N_tT}$, and to noise $\bd{n}$. In the iterative
detection and decoding scenario, soft information about $\bd{x}$ can be fed back from an
FEC decoder (if present) or from a function that computes extrinsic soft information on lattice points (e.g.,  BP in Fig.~\ref{icmmse}); it is made available to the filter bank in the form of
probabilities of valid realizations of transmitted lattice vector $\bd{x}$,
or its elements $x_i$; i.e., either at the vector level $\bd{x}$,
$\{\Prob(\bd{x} = \phi(\bd{C}^{\rm T}))\left|\phi(\bd{C}^{\rm T} \in
\mathcal{C}(\Lambda, \bd{u}_0, \mathcal{R}) \right. \}$, or at the
coordinate level---e.g.\ in the case when coordinate interleaving
\cite{ion:int} is used to scramble the coordinates of several
vectors $\bd{x}$ prior to transmission. In the latter case the
structure present in  the different multidimensional lattice  points
is destroyed during transmission through the channel; not only does
this mean that the coordinate probabilities supplied by the decoder
have to be unscrambled before being fed back to the IC-MMSE filter
for IC, see Fig.~\ref{icmmse}, but the
performance can be improved (over the non-interleaved scenario) even
in an uncoded system (see Section~\ref{itercoord}). Let $\bd{x}_{\rm e}$ denote the soft estimate of $\bd{x}$ ($\bd{x}_{\rm e}
= \bd{0}$ in the first iteration). An iterative receiver aims at iteratively canceling the interference  prior to
filtering by forming a soft interference estimator in two ways:
\begin{enumerate}
\item {\em Vector level  feedback}:    
\be \textstyle \!\!\!\!\bd{x}_{\rm e} =
\sum_{\phi({\scriptsize \bd{C}}^{\rm T}) \in \mathcal{C}(\Lambda, \bd{u}_0, \mathcal{R}) }
\phi(\bd{C}^{\rm T}) \Prob\left(\bd{x} = \phi(\bd{C}^{\rm T})\right)\ee 
\item{\em Coordinate level  feedback}:
If ${\cal K}_i$ is the $i$-th coordinate's alphabet,  the average interference value at position $i$ is
\be \textstyle {x}_{{\rm e}}^{i} =
\sum_{\zeta \in {\cal K}_i}
\zeta \Prob({x}_i = \zeta). \ee
\end{enumerate}
Denote by $\bd{x}^{\overline{i}}_{\rm e}$, $i\!
\!=\!\!1,\ldots\!,m$,  the vector obtained by setting the $i$-th
element of $\bd{x}_{\rm e}$ to zero, i.e., $\bd{x}^{\overline{i}}_{\rm e} =
[\cdots,x^{i-1}_{e},0,x^{i+1}_{e},\cdots]^{\rm T}$; the output
of the IC-MMSE filter bank's $i$-th branch, $\bd{m}_i$, is \be
\label{e:ICMMSE} \hat{x}_i = \bd{m}^{\rm T}_i (\bd{y} -
\bd{H}\bd{x}^{\overline{i}}_e),\ee 
where $\bd{m}_i$ is subject to the unit power constraint \be
\label{e:powercontraint}{\bd{m}_i}^{\rm T} \bd{h}_i = 1 .\ee
Following \cite{xiaodong:mmse}, \cite[Sec.\ 6.10.1]{far:ada} the IC-MMSE filter bank
$\{\bd{m}_{i}\}$ and the $i$-th branch's MSE $\sigma^2_i \!\!=\!\! {\rm E}\!\left\{||x_i \!-\!
\hat{x}_i||^2 \!\right\} $   are\footnote{The coordinates of $\bd{x}$ are assumed uncorrelated.} 
\be \bd{m}_i  =  \bd{m}^c_i + \frac{1-\bd{h}_i^{{\rm
T}}\bd{m}^c_i}{\bd{h}^T_i
\bd{R}^{-1}_i\bd{h}_i}\bd{R}^{-1}_i\bd{h}_i, \label{e:MMSE} \ee
\ba
\sigma^2_i & = &({P}/{2N_t}) - (\bd{m}^c_i)^{{\rm
T}}\bd{R}_i\bd{m}^c_i + \frac{1-\bd{h}_i^{{\rm
T}}\bd{m}^c_i}{{\bd{h}_i}^{\rm
T}\bd{R}^{-1}_i\bd{h}_i},\label{e:MSE}\\ 
\bd{m}^c_i & = &({P}/{2N_t})\bd{R}^{-1}_i \bd{h}_i, \\
\bd{R}_i & = &\bd{H}\bd{Q}_i\bd{H}^{\rm H} + ({N_0}/{2}) \bd{I}, \label{e:IC} \\
\bd{Q}_i &  = & ({P}/{2N_t})\bd{I} - \mbox{diag}\{ \bd{x}^{\overline{i}}_{\rm e} \} \mbox{diag}\{ \bd{x}^{\overline{i}}_{\rm e} \}.
\ea
The   $i$-th element's soft estimate post-IC-MMSE filtering is
\be \hat{x}_i = x_i +
\hat{n}_i, \label{xhati}\ee 
with $\hat{n}_i \sim \mathcal{N}(0, \sigma^2_i),$ or
 in matrix form as $ \hat{\bd{x}} = \bd{x} + \hat{\bd{n}}.$
The non-iterative case \cite[Sec.\ 6.10.1]{far:ada} corresponds to $\bd{x}^{\overline{i}}_{\rm e}\!\!=\!\bd{0}$ above.

\subsection{BP Detector for Lattice Code Based on TG Representation}\label{TG}
After IC-MMSE equalization, the soft estimate $\bd{\hat{x}}$ of a lattice point
 is obtained.  Recall that in lattice space-time
schemes, the codebook of transmitted vectors $\bd{x}$ is a lattice
code $\mathcal{C}(\Lambda,\bd{u}_0,\mathcal{R})$, where the
generator matrix of $\Lambda$ is $\bd{\Gamma}\bd{B}$. For 
 simplicity,  bet  $\bd{B}$ be a generic lattice generator 
matrix. Lattice detection is to either
decide which lattice point {\it inside the shaping region} has the
minimum distance to $\bd{\hat{x}}$, or calculate the soft
information (e.g., in the form of probability or log-likelihood
ratio) about  each candidate lattice point. The first detection
criterion leads to hard decision detectors---e.g., maximum likelihood (ML). The second decoding criterion leads to soft decision detectors,
which can be used in iterations between detection and decoding. In this
section, a novel TG based lattice decoding algorithm is
introduced. For  simplicity, assume
  an $m$-dimensional lattice code, i.e., $\bd{\hat{x}} \in
\mathbb{R}^{m}$.

The novel lattice decoding algorithm introduced below relies on TG
representations of lattices \cite{ban:tan}, which are enabled by lattice partitioning; all  lattice points (those inside
the shaping region are of interest) are partitioned into several subgroups (cosets). Each subgroup
includes several different lattice points, and is labeled by a
well-defined Abelian group block codeword. Then, a reduced-complexity
soft-output lattice detector  can be obtained by operating on the smaller
number of cosets instead of lattice points. The labels of all
cosets form an Abelian block code, which  can be
represented by a TG similar to LDPC codes. BP on a lattice is performed on its non-binary label TG to
yield  total and extrinsic APPs of the labels and their coordinates, as described in the sequel. The APPs of individual lattice points are obtained in a final step described in Section~\ref{pointAPP}. 

A somewhat subtler point is that lattice  partitioning revolves around an orthogonal sublattice $\Lambda^\prime$ of $\Lambda$, and the quotient group $\Lambda/\Lambda^\prime$; $|\Lambda/\Lambda^\prime|$ is finite iff $\Lambda$ and $\Lambda^\prime$ have equal dimensionalities. The most straightforward way of obtaining $\Lambda^\prime$ is by Gram-Schmidt (G-S) orthogonalization of $\Lambda$'s generator matrix, whereby all orthogonal G-S directions intercept $\Lambda$, and  the intersection naturally forms a sublattice of the same dimensionality as $\Lambda$; alternatively, the orthogonal sublattice will have to be obtained by  means other than G-S orthogonalization.

\subsubsection{Gram-Schmidt Orthogonalization} Given a generator matrix
    $\bd{B}=\left[ \bd{b}_1 {\ldots} \bd{b}_m\right]$, obtain $m$ orthogonal vectors 
     $\{\! \bd{w}_i\!\}^{m}_{i{=}1}$, $\bd{W}{=}\left[ \bd{w}_1 {\ldots} \bd{w}_m\right]$.\footnote{$\bd{w}_1{=}\bd{b}_1$,  $\bd{w}_i{=}\bd{b}_i {-}\! \sum^{i-1}_{j=1} \mu_{ij}\bd{w}_j,\ 
         i{=}2,{\ldots},m$, 
  $ \mu_{ij} {\df} {\langle\bd{b}_i,\! \bd{w}_j \!\rangle}/{\langle \bd{w}_j,\!
         \bd{w}_j\!\rangle}$.} 
         Let  $W_i$ be the vector
         space spanned by $\bd{w}_i$, $W_i {=} \alpha \bd{w}_i, \alpha {\in} \mathbb{R}$; ${\cal W}{\df} \{W_i\}_{i=1}^m$ is a coordinate system.

\subsubsection{Lattice Label Groups $G_i$} 
    Let  $P_{W_i}(\Lambda)$ be the projection of $\Lambda$ onto the vector space $W_i$, and $\Lambda_{W_i} {\df} \Lambda {\cap}
    W_i$.  The quotient group $P_{W_i}(\Lambda)/\Lambda_{W_i}$ is called a label group $G_i$; $\Lambda$ is  partitioned into a finite set of cosets labeled by $m$-tuples from $G\doteq G_1 \times \ldots {\times} G_m$.
The finite set of all label $m$-tuples, denoted $\bd{L}(\Lambda)$, is called the label code, and uses $G$ as its alphabet space \cite{ban:tan}.

\subsubsection{Lattice Label Code $\bd{L}(\Lambda)$} Due to the isomorphism $G_i \cong {\mathbb Z}_{g_i}$, with $g_i \df |{G}_i|$, let ${G} = \mathbb{Z}_{g_1} \times \cdots \times \mathbb{Z}_{g_m}$.
A lattice point will be labeled by the label of the coset to which it belongs. 
The label code $\bd{L}(\Lambda)$ is an Abelian
    block code. Let $\bd{l} = [l_1 \ldots l_m]^{\rm T}$ denote a label, and $\Lambda(\bd{l})$ denote the set of lattice points
    sharing the label  $\bd{l}$; clearly, labeling is invariant to translations of $\Lambda$ by $\bd{u}_0$. Let $\bd{L}(\Lambda)$,
    $\bd{L}(\mathcal{C}(\Lambda,\bd{u}_0,\mathcal{R}))$  denote the label
    codes of $\Lambda$, respectively of the subset of translated lattice points inside a
    shaping region $\mathcal{R}$. Then, a translated lattice point inside $\mathcal{R}$ will have a label $\bd{l} \in
    \bd{L}(\mathcal{C}(\Lambda,\bd{u}_0,\mathcal{R}))$.

\subsubsection{Finding a Set of Generator Vectors $ {\cal {V}^\ast} \df \{ \bd{v}^{\ast}_i\}^{n}_{i=1}$ for the  dual label code $\bd{L}(\Lambda)^{\ast}$ of $\Lambda$'s  label code $\bd{L}(\Lambda)$ \protect\mbox{\rm \cite{ban:tan}}} 
The generator vectors $\{\bd{v}^{\ast}_i\}^{n}_{i=1}$   characterize the lattice $\Lambda$ like a parity check equation characterizes  a  linear block code, and have the following property: all  labels in $\bd{L}(\Lambda)$ are orthogonal to every vector $\bd{v}_i$ in $ \{ \bd{v}^{\ast}_i\}^{n}_{i=1}$, i.e., 

\be \textstyle \label{e:check} {\bd{v}^{\ast}_i}^{\rm T} \bd{L}(\Lambda) = 0~\mbox{\rm mod}~\lcm(g_1,g_2,\cdots,g_m) \ee 
where $\lcm(\cdot,\ldots,\cdot)$ is the least common multiple.

\subsubsection{Lattice (Label) Tanner Graph} The generator vectors $\{
    \bd{v}^{\ast}_i\}^{n}_{i=1}$ act as check equations for the label code $\bd{L}(\Lambda)$, according to
    (\ref{e:check}). Each coordinate of a label  $\bd l$
    corresponds to a variable node, and each generator vector that defines a
    check equation involving several label coordinates  corresponds to a check node.
    A TG is constructed according to the constraints placed on  label coordinates by the generator vectors $\{
    \bd{v}^{\ast}_i\}^{n}_{i=1}$. In general, the check equations
     are not over $GF(2)$, unless the cardinalities of the label groups $\bd{G}_i$ are all
    two. Thereby, the TG of a lattice is, generally, non-binary.
\newtheorem{ex3}[ex2]{Example}
\begin{ex3}
\label{ex3}(checkerboard lattice, $\Lambda \!= \!D_4\!\in \!{\mathbb R}^4$)
The matrix
\be \label{e:D4} \bd{B} = \left[ \renewcommand{\arraystretch}{.7}\begin{array}{cccc} 1 & 1 & 1 & 2 \\
                               1 & 0 & 1 & 0  \\
                               0 & 1 & 1 & 0 \\
                               0 & 0 & 1 & 0  \end{array}\renewcommand{\arraystretch}{1} \right] \ee
is a generator  for $D_4$, with associated G-S vectors 
\renewcommand{\arraystretch}{.5}\ba
\bd{w}_1 & = & [\begin{array}{cccc} 1, & 1, & 0, & 0 \end{array}]^{\rm T} \nonumber \\
\bd{w}_2 & = & [\begin{array}{cccc} 1/2, & -1/2, & 1, & 0 \end{array} ]^{\rm T} \nonumber \\
\bd{w}_3 & = & [\begin{array}{cccc} -1/3, & 1/3, & 1/3, & 1 \end{array}]^{\rm T} \nonumber \\
\bd{w}_4 & = & [\begin{array}{cccc} 1/2, & -1/2, & -1/2, & 1/2
\end{array}]^{\rm T}
\ea  \renewcommand{\arraystretch}{1}\noindent In the coordinate system $\{W_i\}^{4}_{i=1} =
{\rm span\{\bd{w}_i\}}^4_{i=1}$  the following projections $P_{W_i}(\Lambda)$
and cross-sections $\Lambda_{W_i}$ are obtained: 
\ba \textstyle P_{W_1}(\Lambda) =
\frac{\mathbb{Z}}{\sqrt{2}}\frac{\bd{w}_1}{|| \bd{w}_1||} &~&\textstyle 
\Lambda_{W_1} =
\sqrt{2}\mathbb{Z} \frac{\bd{w}_1}{|| \bd{w}_1||} \nonumber \\
\textstyle P_{W_2}(\Lambda) = \frac{\mathbb{Z}}{\sqrt{6}}\frac{\bd{w}_2}{||
\bd{w}_2||} &~&\textstyle  \Lambda_{W_2} = \sqrt{6}\mathbb{Z}
\frac{\bd{w}_2}{|| \bd{w}_2||}
\nonumber \\
\textstyle P_{W_3}(\Lambda) = \frac{\mathbb{Z}}{\sqrt{3}}\frac{\bd{w}_3}{||
\bd{w}_3||} &~&\textstyle  \Lambda_{W_3} = 2\sqrt{3}\mathbb{Z}
\frac{\bd{w}_3}{||
\bd{w}_3||} \nonumber \\
\textstyle P_{W_4}(\Lambda) = \mathbb{Z}\frac{\bd{w}_4}{|| \bd{w}_4||} &~&\textstyle 
\Lambda_{W_4} = 2\mathbb{Z} \frac{\bd{w}_4}{|| \bd{w}_4 ||}
\nonumber.  
\ea 
This results in the following quotient groups for
$D_4$:
$G_1(\Lambda) = \left\{ 0, \frac{\sqrt{2}}{2}\right\}$, 
$G_2(\Lambda) = \left\{ 0, \frac{\sqrt{6}}{6}, \frac{\sqrt{6}}{3},\frac{\sqrt{6}}{2},\frac{2\sqrt{6}}{3},\frac{5\sqrt{6}}{6} \right\}$,
$G_3(\Lambda) = \left\{ 0, \frac{\sqrt{3}}{3}, \frac{2\sqrt{3}}{3},\sqrt{3},\frac{4\sqrt{3}}{3},\frac{5\sqrt{3}}{3} \right\}$,
$G_4(\Lambda) {=} \left\{ 0, 1 \right\}$.
The label code   $\bd{L}(\Lambda)$ and its dual $\bd{L}^{\ast}(\Lambda) {\subset} \mathbb{Z}_2 {\times}
\mathbb{Z}_6 {\times} \mathbb{Z}_6 {\times} \mathbb{Z}_2 $ are, respectively, 
\ba  
\begin{array}{ccccccc} \bd{L}(\Lambda) & =\left\{ 0000,\right.& 0031,&
0220,& 0251, & 1300,& 1331, \nonumber \\
 & 1520,& 1551, & 1140,& 1111 & 0440,& \left.0411\right\},
\end{array} \\   
\begin{array}{ccccccc} \bd{L}(\Lambda)^\ast & =\left\{ 0000,\right.& 0240,&
0420,& 1511, & 1300,& 1331, \nonumber \\
 & 0451,& 1540, & 1151,& 0031 & 1120,& \left.0211\right\}
\end{array} 
\ea
\cite{ban:tan}. The generator  set for
$\bd{L}(\Lambda)^{\ast}$ is ${\cal V}^{\ast} {=} \{ 1151,0240,0031 \}$. Since $\lcm(g_1,g_2,g_3,g_4) =  6$, the TG of label
code $\bd{L}(\Lambda)$ can be constructed accordingly, as given in
Fig.~\ref{f:TG}, where $v_j$ is the $j$-th check node, and $l_i$
is the $i$-th variable. 
\begin{figure}[bt]
\scalebox{0.7} {\input{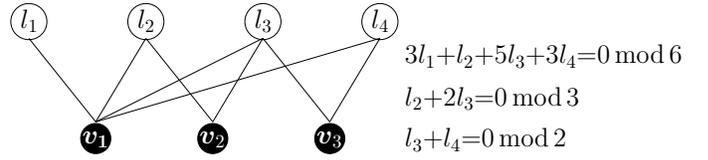}} \caption{Tanner graph (not minimal)
for  $D_4$ of Example~\protect\ref{ex3}, following \protect\cite{ban:tan}; relevant to decoding the 8-D lattice \protect$D_4{\oplus} D_4{\in}\mathbb{R}^8$ and the  codes of Example~\protect\ref{ex2}.}\label{f:TG} 
\end{figure}
The variable nodes
associated with generator vector  $\bd{v}_j^\ast$ are connected to $v_j$; e.g., check node $v_1$ is connected to all four variable nodes,
because all variable nodes are involved
in the first check equation.\hfill $\square$
\end{ex3}

\subsubsection{Non-Binary BP {\rm \cite{Low:Mat}} on a Lattice TG} \label{nonbinarybp}
$P_{W_i}(\bd{\hat{x}})$ denotes the  projection of $\bd{\hat{x}}$, which may not be in $\Lambda$, onto vector
space $W_i$, i.e.\ $P_{W_i}(\bd{\hat{x}}){=}{\hat{\bd{x}}^{{\rm T}}\bd{w}_1}/{|| \bd{w}_1||}$.
In the lattice TG a value  $\alpha \in
\{0,\ldots,g_i{-}1\}$ of the variable node $l_i$ is associated with the hypothesis that $\hat{\bd{x}}$ is an observation of a lattice point whose label has its $i$-th coordinate equal to $\alpha$ (or, whose projection on the vector space
$W_i$ belongs to the coset whose label has $\alpha$ as its $i$-th element);  $\Prob(l_i {=} \alpha)$
is the probability of this hypothesis.

Define messages $q^{\alpha}_{ji}$,
$r^{\alpha}_{ji}$ where  subscripts $i,j$ refer to $i$-th variable
node $l_i$ respectively $j$-th check node $v_j$.
The quantity
$q^{\alpha}_{ji}$ is  the probability of the hypothesis that
$\hat{\bd{x}}$ is an observation of a lattice point whose label has $i$-th coordinate equal to $\alpha$, 
given the information obtained via check nodes other than $v_j$; $r^{\alpha}_{ji}$ is the
probability of  
check $v_j$ being satisfied 
given  that $\hat{\bd{x}}$ is an observation of a lattice point whose label has $i$-th coordinate equal to $\alpha$.
Message passing---adapted from \cite{Low:Mat}---is then:

\be  \textstyle { r^{\alpha}_{ji} = 
\sum_{\sta{\bd{l}{\in} \bd{L}(\Lambda)}{{\bd{v}_j^\ast}^{\rm \scriptsize T}\bd{l}{\equiv} 0}{\ l_i {=} \alpha}}}\ 
\prod_{k {\in} \mathcal{N}(j)\backslash \{i\}}{q^{l_k}_{jk}},
\ee
 \be \textstyle q^{\alpha}_{ji} = K_{ji} f^{\alpha}_i \prod_{k \in \mathcal{M}(i)\backslash 
 \{j\}}{ r^{\alpha}_{ki}}, \ee
where $K_{ji}$ are implicitly defined via $\sum_{\alpha} q^{\alpha}_{ji} {=}
 1$,  $\mathcal{N}(j) $ is the set of variable nodes involved in check equation $v_j$, and $\mathcal{M}(i)$ is the set of check nodes  connected to variable node $l_i$; $f^{\alpha}_{i}$ is the initial probability
 of event $l_i {=} \alpha $ given observation $\hat{\bd{x}}$.
\begin{figure}[b]
\scalebox{1} {\input{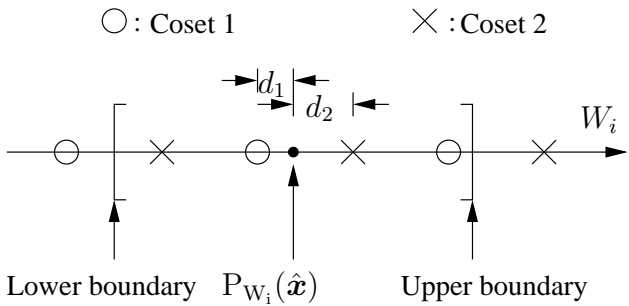}}
\caption{Illustrative projection of a point $\hat{\bd{x}}\in {\mathbb R}^m$ on one of the orthogonal directions $W_i$, $i = 1, \ldots, m$, whose label group  has cardinality $|G_i|=2$.}\label{projection}
\end{figure}  

\subsection{Initializations (in Projection and/or Probability Domains)\label{sec:initlblprob}}
BP requires initializing
$f^{\alpha}_i$  in the {\em labels'} TG. A simple, memoryless SISO module---relating lattice {\em points} to labels in state (or label) space (Sec.\ \ref{pointAPP} and Fig.\ \ref{icmmse})---requires lattice point (log-)likelihoods  for initialization. Most calculations are reusable---between BP and SISO modules, since the orthogonal projectors $\bd{w}_i$ preserve distances; known approximations 
can be employed by judiciously exploiting log-likelihoods, max-log (Jacobi-log) approximations, etc..
Notably, after partitioning  the infinite lattice into finitely many {\em labeled} cosets, not all labels are necessarily used by points in $\cal R$. 

\subsubsection{Projection Domain\label{sec:projdmn}} 
The soft estimate
$\hat{\bd{x}}$ obtained from the LMMSE filters bank is projected onto vector
spaces $\{W_i \}^m_{i=1}$ (Fig.~\ref{projection}); with
$\sigma^2_i$ of \rf{e:MSE}, $f^{\alpha}_i$ is initialized as follows:
\begin{enumerate}[i.]
\item $\forall  \bd{l}\in \bd{L}\left(\mathcal{C}(\Lambda,\bd{u}_0,\mathcal{R})\right)$, find closest $\lambda \in {\cal R} \cap \{\Lambda(\bd{l})+ \bd{u}_0\}$:
\be \textstyle
\!\!\!\bd{\lambda}_{\min}(\bd{l}) {=} \arg \min_{\bd{\lambda} {\in}
  \Lambda(\bd{l})} \sum_{i=1}^m | P_{W_i}(\bd{\hat{x}}) {-}  P_{W_i}(\bd{\lambda})  |^2 \label{mincoorddist}
\ee
\item Calculate 
the probability of (subgroup with) label $\bd{l}$ via 
\be \!\!\!\!\!\!\!\!\Prob(\bd{l}) {=}
\frac{\exp\left(-\sum^m_{i=1}{d^2_{i}\left(\bd{\lambda}_{\min}\left(\bd{l}\right)\right)}/{(2\sigma^2_i)}\right)}
{\sum_{\bd{\ell} {\in} \bd{L}(\mathcal{C}(\Lambda,\bd{u}_0,\mathcal{R}))}
\exp\left(-\sum^m_{i=1}{d^2_{i}\left(\bd{\lambda}_{\min}\left(\bd{\ell}\right)\right)}/{(2\sigma^2_i)}\right)
} \label{prlbl1}
\ee 
with $d_{i}(\bd{\lambda}_{\min}(\bd{l})) {=}
|P_{W_i}(\bd{\hat{x}}) {-} P_{W_i}(\bd{\lambda}_{\min}(\bd{l}))|$.
\end{enumerate}

\newtheorem{rem1}{Remark}
\begin{rem1}
\label{remark1}(Simpler initialization) One can separately test $\hat{\bd{x}}$
along  each  ${W_i}$ in isolation from others---i.e., no precaution  to verify that selecting the closest projection coordinate in each direction  ${W_i}$ aggregates to a lattice point. Let ${\cal S}_i(\bd{\ell})\doteq \left\{ x \left| x=\bd{\lambda}^{\rm T} {\bd{w}_i}/{\|\bd{w}_i\|}, \forall \bd{\lambda}{\in} \Lambda(\bd{\ell}) {\cap} {\cal R}   \right. \right\}$ be the 
projections onto  ${\bd{w}_i}/{\|\bd{w}_i\|}$ of all lattice points in $\cal R$ that have label $\bd{\ell}$;  
 $\forall\bd{l}$: 
\begin{enumerate}[i.]
\item Find the closest coordinate projection along $W_i$ 
\be  
 t_{i,\min}(\bd{l}) {=}  \textstyle \arg \min_{t\in {\cal S}_i(\bd{l})}|t{-}P_{W_i}(\bd{\hat{x}}| \label{mincoorddistS}
\ee
\item  Calculate the probability of (subgroup with) label $\bd{l}$ via
\be \!\!\!\!\!\!\!\!\!\!\!\!\!\Prob(\bd{l}) = \frac{\exp\left(-\sum^m_{i=1}{{d_i^\prime}^2\left(t_{i,\min}\left(\bd{l}\right)\right)}/{(2\sigma^2_i)}\right)}{\sum_{\bd{\ell} {\in} \bd{L}(\mathcal{C}(\Lambda,\bd{u}_0,\mathcal{R}))}  \exp\left(-\!\sum^m_{i=1}{{d_i^\prime}^2\left(t_{i,\min}\left(\bd{\ell}\right)\right)}/{(2\sigma^2_i)}\right)  } \label{prlbl2}
\ee 
with  $d_i^{\prime}(\cdot){=}\left|\cdot-P_{W_i}(\bd{\hat{x}})\right|$.
\hfill $\square$
\end{enumerate}
Lastly, $f^{\alpha}_{i}$ is initialized according to 
\be \textstyle
\label{e:ini} f^{\alpha}_i = \sum_{\scriptsize \bd{l}, l_i = \alpha}
\Prob(\bd{l})
\ee
Thereafter, $q^{\alpha}_{ji} $ is initialized to $f^{\alpha}_{i}$,  
and BP 
updates $r^{\alpha}_{ji}$ and
$q^{\alpha}_{ji}$ iteratively 
up to a predetermined number of iterations. 
\end{rem1}

\subsubsection{Probability Domain}
The likelihoods of each valid coordinate\footnote{A {\em real} coordinate of a lattice point, not an {\em integer} coordinate of a label.} value for $\bd{x}{\in} \Lambda$ at the $k$-th MIMO channel use can be calculated from the soft estimates in $\hat{\bd{x}}$.\footnote{The subscript $k$, which would indicate the time index of the relevant MIMO channel use, is omitted here and in Fig.~\protect\ref{icmmse} for simplicity of notation.} With $K$ a constant,
\be P(\hat{x}_i | x_i = c^j) = K \exp
\left(-{||\hat{x}_i - c^j||^2}/{2\sigma^2_i} \right),\ee
where $c^j$ is  the $j$-th valid value of the real $i$-th  coordinate $x_i$ of $\bd{x} {\in} \Lambda \cap {\cal R}$. Then,
the likelihood of each valid value of ${x}_i$ at $k$-th MIMO channel use will supply the component
$P_k(c^j;{\rm I})$ of a vector input $P_k(\bd{c};{\rm I})$ to a SISO APP module (see below), following the model and notations in \cite{BenedettoDMP:97}; as in \cite{BenedettoDMP:97}, $C_k^j$ will denote a random process  enacted by a sequence of (coordinate) symbols taking values  in some alphabet $\{c^j|j\in {\cal J}\}$---which, nonetheless, may be non-binary, i.e.\ 
$|{\cal J}|\geq2$.

\subsection{Extrinsic APP---either Point- or Coordinate-Wise---post BP} \label{pointAPP}
\begin{figure}[bth]
\scalebox{0.8} {\input{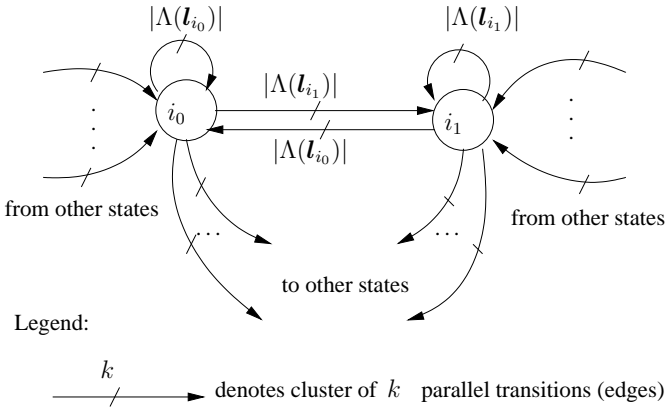}} \caption{State transition diagram for Markov process representing a sequence of lattice points. Edges occur in clusters because every label generally covers more than one point in the shaping region. States are label indices; the state at any time is the {\em index} of the label that contains the most recent lattice point output by the Markov source. When the Markov source outputs a new point it transitions into the state indexing the label that contains the new point.\label{markov}}
\end{figure}

\begin{figure*}[bt]
\scalebox{0.86} {\input{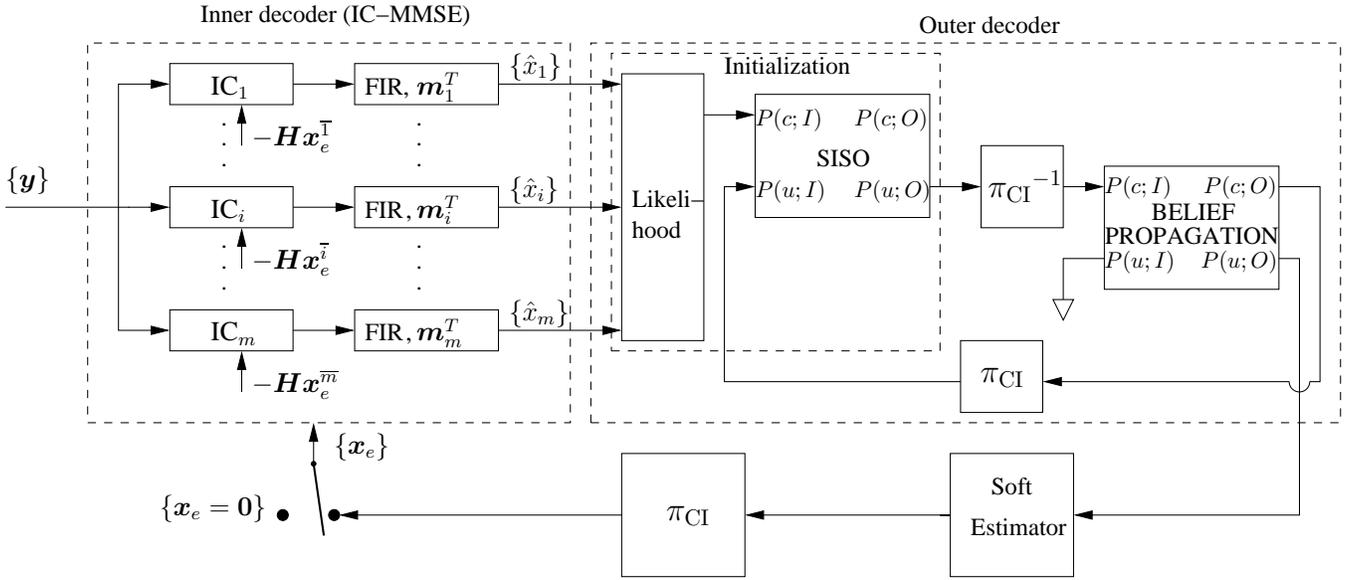}} 
\caption{Generic 
iterative receiver for decoding a lattice on its TG, with  BP on the lattice performed in the presence of coordinate interleaving, and extrinsic soft information on lattice coordinates; uncoded case---lattice constellation employed \protect{\em per se}, without additional coding gain derived from FEC encoding.}
\label{icmmse}
\end{figure*}

In order to implement iterative receivers it is necessary to compute the a posteriori probability at the end of BP.
After the last iteration, the BP returns
$r^{\alpha}_{ji}$ and $q^{\alpha}_{ji},~\forall \alpha,i,j.$ Then,
the total a posteriori  probability  $\Prob(l_i = \alpha)$ is computed as 
\be \textstyle \Prob(l_i
= \alpha) = f^{\alpha}_{i} \prod_{j \in \mathcal{M}(i)}
r^{\alpha}_{ji}, \ee 
and the total a posteriori probability of each label is
given by 
\be \textstyle \label{e:totalAPP}\Prob(\bd{l} = \{\alpha_1, \alpha_2,\cdots, \alpha_m\}) =
\prod^{m}_{i=1} \Prob(l_i = \alpha_i).~ \ee
In Appendix~\ref{app1} it is shown that
when a lattice is represented by a TG,    the Markov process in Fig.~\ref{markov} can be associated with the model for soft detection of lattice points;
also, that
the extrinsic APPs $P^{\rm BP}_k(c^j;{\rm O})$, $P^{\rm BP}_k(u^j;{\rm O})$ after BP, corresponding to the $k$-th transition between states, can be computed as:
\ba \textstyle P^{\rm BP}_k(c^j;{\rm O})&= & \textstyle \sum_{e:C_k^j(e) = c^j}
\Prob \left( \bd{l}_{s^S(e)}\right)\prod^m_{i=1}P_k[u^i(e);{\rm I}] \nonumber \\
& & \textstyle \times \prod^m_{i=1;i \neq j} P_k[c^i(e);{\rm I}], \label{pcotg}\ea
 \ba P^{\rm BP}_k(u^j;{\rm O}) &\!\!\!=\!\!\!\! & \textstyle \sum_{e:U_k^j(e) = u^j}
\Prob \left( \bd{l}_{s^S(e)}\right)\prod^m_{i=1;i \neq j}P_k[u^i(e);{\rm I}] \nonumber \\
& & \textstyle \times \prod^m_{i=1} P_k[c^i(e);{\rm I}], \label{puotg} \ea
 where $\bd{l}_{s^S(e)}$ is the label indexed by the integer value of the starting state $s^S(e)$ of
edge $e$.  $P_k[u^i(e);{\rm I}]$, $P_k(c^i(e);{\rm I})$ are the a priori probabilities of an unencoded,
respectively encoded, symbol element (in this case a coordinate\footnote{I.e., not necessarily a binary symbol, or bit.}) at position $i$, which are associated with edge $e$ \cite{BenedettoDMP:97}. In a serial concatenation such as in Fig.~\ref{icmmse}, the
unencoded symbol elements are assumed to be identically distributed according to a uniform distribution, and thereby
$P_k(u^i(e);{\rm I})$ equal the reciprocal of the alphabet size at position $i$. $P_k(c^i(e);{\rm I})$ are the likelihoods of lattice point coordinates, computable as in the TG initialization step.

In the sense of \cite{mud:min}--\cite{laf:asy}, the trellis in Fig.\ \ref{markov} is {\em proper}, {\em one-to-one}, but {\em nonlinear} (w.r.t.\  a finite subset of points carved from the infinite lattice); the edge label alphabet is the real  alphabet that describes the real and imaginary coordinates of the skewed lattice. 

\newtheorem{ex5}[ex2]{Example}
\begin{ex5}
\label{ex5}
For the super-orthogonal lattice code from $\mathbb{R}^8$ (see Example~\ref{ex2}) the size of the trellis edge label alphabet  is $q=3$, because the nonzero realizations of $\bd{\chi}$, $\bd{\chi}^\prime$ are all of the sixteen 4-dimensional vectors with elements $\pm 1$, and there are four null coordinates.  
However, when restricted to each direct sum lattice term $D_4$ the edge alphabet is binary ($\pm1$).
\hfill $\square$
\end{ex5}

\section{Application to the detection of a $2{\times}2$ MIMO super-orthogonal lattice space-time code}
The 
algorithm devised in Sec.~\ref{sec:detector} is illustrated, in combination with 
hypothesis testing, on 
the size-32, super-orthogonal, space-time constellation 
identified in Example~\ref{ex2} as (carved from) a lattice in $\mathbb{R}^8$ \cite{IonescuMYL:01,ion:spa}, \cite{siw:impcon}--\cite{ion:fad}.
Per Sec.\ \ref{sec:stlattcd} notation, this lattice code is $\mathcal{C}(D_4\oplus D_4, \bd{0},
\mathcal{R}\oplus\mathcal{R})\in \mathbb{R}^8$, in short $\mathcal{C}$, where $\mathcal{R}$ is a shaping region that carves out the set $\mathcal{D}$ (within second shell of $D_4$, see Example~\ref{ex2}) from the whole of $D_4$. One of the two most useful 4-D lattices, $D_4$ has rich structure, and is iso-dual \cite[Sec.\ 7.2]{con:sph} with gain $\gamma(D_4)=1.414$---i.e., within 3\% of the maximum gain in 4-D (Hermite's constant).

\newtheorem{rem2}[rem1]{Remark}
\begin{rem2}
\label{remark2}
If $\mathcal{C}$ is combined with a channel code, or used in 
trellis coded modulation---natural scenarios if operating  near capacity is a goal---then it is the FEC code's redundancy (memory) that ultimately determines whether one or the other direct-sum $D_4$ terms is selected during space-time 
modulation (as in the trellis space-time  codes from \cite{IonescuMYL:01,ion:spa,jaf:sup}). The  illustrative example below does not rely on an actual channel code, since FEC codes are relatively well-understood, while the focus  herein is on the lattice search. Adding a channel code would be an otherwise straightforward effort, and the extrinsic soft outputs in Fig.~\ref{icmmse} are precisely meant for enabling such purpose---with flexibility for iterating between FEC  decoder(s) (possibly concatenated) and (MIMO) detectors. Herein, the 
$\log_2 |\mathcal{C}|=5$ bits to be simultaneously mapped to $\mathcal{C}$ are parsed in 1- and 4-bit subsets; the former  selects one or the other $D_4$ lattices, independently of the remaining four bits; thus, absent FEC, the latter (four) bits do not carry information about the fifth bit, which can be demodulated separately. As it turns out, but not dwelled upon herein, the latter bits do convey extrinsic information about each other due to the structure of lattice points in $\mathcal{C}$; see Example~\ref{ex2}.\hfill $\square$

\end{rem2}

\subsection{Receiver for Quasistatic Scenarios}\label{quasistaticrx}
Consider the superorthogonal space-time
code given in Example \ref{ex2}. The ML
receiver for $\bd{x}_{\oplus}$ is given by 
\be \textstyle \bd{x}_{\oplus; \protect\mbox{\rm   \scriptsize ML}}
= \arg \min_{\bd{x}_{\oplus}}
  ||\bd{y} - \bd{H}_{\oplus}\bd{x}_{\oplus}||^2. \ee The ML receiver is usually
computationally complicated since it needs to examine  all
valid lattice points (complexity grows exponentially). The  algorithm
introduced in Section~{\ref{sec:detector}} offers a computationally  efficient solution.

Recall that for a  superorthogonal space-time code (see Example~\ref{ex2}), either all $\chi_l$ or all $\chi_l^\prime$ are zeros, which
identifies two  hypotheses:  hypothesis ${\cal H}_1$ is that
$\chi_l^\prime$ vanish, and the basis matrices $\bd{C}$ are
chosen; hypothesis ${\cal H}_2$ is that ${\chi_l}$ vanish, and 
basis matrices $\bd{C}^\prime$ are chosen.
When hypotheses ${\cal H}_1$, ${\cal H}_2$ are true,
the transmission model \rf{e:superothogonal} becomes, respectively,
\be \bd{y} {=} \begin{cases} \bd{H}^1_{\oplus} \,\bd{\chi} + \bd{n},& {\cal H}_1\ {\rm true}\\\bd{H}^2_{\oplus} \,\bd{\chi}^\prime + \bd{n},& {\cal H}_2 \ {\rm true}\end{cases}\label{h1or2}\ee
Due to
the orthogonality of matrices $\bd{H}^k_{\oplus},~k=1,2,$ the MMSE
filters for $\bd{\chi}$, $\bd{\chi}^\prime$ are the corresponding matched filters 
\be \bd{M}^k =
{\alpha^{-1}}{ \bd{H}^k_{\oplus} }^{\rm H},~k = 1,2
\ee 
where $\bd{M}^k$ is the MMSE filter for hypothesis ${\cal H}_k$. The
outputs of MMSE filters for hypotheses ${\cal H}_1$, ${\cal H}_2$ become 
\ba \hat{\bd{\chi}} & = & \bd{M}^1\bd{y} ={\alpha^{-1}}{ \bd{H}^1_{\oplus} }^{\rm H} \bd{y} = \bd{\chi} + \bd{\tilde{n}}^1 \label{e:chi1}\\
\widehat{\bd{\chi}^\prime} & = & \bd{M}^2\bd{y} = {\alpha^{-1}}{ \bd{H}^2_{\oplus} }^{\rm H} \bd{y}  {=} \bd{\chi}^\prime +
\bd{\tilde{n}}^2 \label{e:chi2} \ea 
where $\bd{\tilde{n}}^1$ and
$\bd{\tilde{n}}^2$ are estimation noise after filtering for
hypothesis ${\cal H}_1$ respectively ${\cal H}_2$. It is not difficult to
see that $\bd{\tilde{n}}^k, k{=} 1,2 $ are white multivariate Gaussian
random vectors, i.e.\ $\bd{\tilde{n}}^k {\sim}
\mathcal{N}(0,\frac{N_0}{2\alpha}\bd{I})$. Note that IC is unnecessary
is this scenario, and the estimates of \rf{e:chi1}, \rf{e:chi2}
are interference-free estimates of $\bd{\chi}$ respectively $\bd{\chi}^\prime$, due to  orthogonality of $\bd{H}^k_{\oplus}$.
The {\em a posteriori} probability of ${\cal H}_1$ given $\bd{y}$ is (in log-domain): 
\be \textstyle \log\left(\Prob({\cal H}_1 | \bd{y})\right)  = \log\left(\sum_{\bd{\chi}} \Prob({\cal H}_1,\bd{\chi}|\bd{y})\right) \label{e:ProHypo}. \ee
Summing over all valid $\bd{\chi}$ patterns in \rf{e:ProHypo} becomes infeasible
as the length of $\bd{\chi}$ increases. 
One can re-write \rf{e:ProHypo} using the well-known max-log approximation $\log\sum_j a_j\approx \max_j\log a_j$ (or, the more accurate Jac-log approximation, see, e.g., \cite{Hochwald:LSD}) 
\be \textstyle \Prob({\cal H}_1 |
\bd{y})  \approx  \max_{\bd{\chi}} \Prob({\cal H}_1,\bd{\chi}|\bd{y}) \sim  p(\bd{y} |\bd{H}^1_{\oplus}, \bd{\chi}_{max}) \label{aposth1maxlog}\ee 
(back in probability domain, for convenience) with
\ba \textstyle  \bd{\chi}_{\max} & {=} & \textstyle \arg \max_{\bd{\chi}} p\left(\bd{y} \left|
\bd{H}^1_{\oplus}, \bd{\chi}\right.\right)  {=}\arg \min_{\bd{\chi}} \left\|\bd{y} {-} \bd{H}^1_{\oplus} \bd{\chi}\right\|^2 \nonumber \\
 & {=} & \arg \min_{\bd{\chi}} \left\|{\bd{H}_\oplus^1}^{\rm H}\left(\bd{y} {-} \bd{H}^1_{\oplus} \bd{\chi}\right)\right\|^2\nonumber\\
& = &  \textstyle \arg \min_{\bd{\chi}} \alpha^2\|\hat{\bd{\chi}} - \bd{\chi} \|^2  {=}  \mbox{sign}~(\hat{\bd{\chi}}) \label{e:H1}\ea 
where
$\hat{\bd{\chi}}$ is the 
${\cal H}_1$ hypothesis statistic of \rf{e:chi1}. \rf{e:H1} used the facts that $\bd{H}^1_{\oplus}$ is unitary (up to a scalar, see \rf{Hoplusunitary}), and that the nonzero realizations of $\bd{\chi}$, $\bd{\chi}^\prime$ are all of the sixteen 4-dimensional vectors with elements $\pm 1$ (Example~\ref{ex2}). Similarly,  
\be \textstyle \Prob\left({\cal H}_2|
\bd{y}\right) \approx \max_{\bd{\chi}^\prime} \Prob\left(\left.{\cal H}_2,\bd{\chi}^\prime\right|\bd{y}\right) \sim
p(\bd{\bd{y}}| \bd{H}^2_{\oplus}, \bd{\chi}^\prime_{\max}) \label{e:H2}
\ee 
\be \textstyle \bd{\chi}^\prime_{\max} \df  \arg \min_{\bd{\chi}^\prime} \|\bd{y} -
\bd{H}^2_{\oplus} \bd{\chi}^\prime\|^2 = \mbox{sign}~(\widehat{\bd{\chi}^\prime}).
\ee 
The log-likelihood hypotheses ratio is (after the max-log approximations \rf{aposth1maxlog}, \rf{e:H2}, see \cite[Sec.\ II.C]{Hochwald:LSD}) 
\ba L({\cal H}) & = &
\log\frac{\Prob({\cal H}_1|\bd{y})}{\Prob({\cal H}_2|\bd{y})}  \approx  \log\frac{p(\bd{\bd{y}}| \bd{H}^1_{\oplus},
\bd{\chi}_{\max})}{p(\bd{\bd{y}}| \bd{H}^2_{\oplus},
\bd{\chi}^\prime_{\max})} \nonumber \\
& = & \frac{2 \alpha}{N_0} \left( || \bd{y} - \bd{H}^2_{\oplus}
\bd{\chi}^\prime_{\max}||^2 -  || \bd{y} - \bd{H}^1_{\oplus}
\bd{\chi}_{\max}||^2\right) \nonumber  \\
& = & \frac{4 \alpha}{N_0} \left(
\bd{y}^{\rm H}\bd{H}^2_{\oplus}\bd{\chi}_{\max} -
\bd{y}^{\rm H}\bd{H}^1_{\oplus}\bd{\chi}^\prime_{\max} \right) \nonumber
\\
& = & \frac{4 \alpha^2}{N_0} \left(
\hat{\bd{\chi}}^{\rm H}\bd{\chi}_{\max} -
\widehat{\bd{\chi}^\prime}^{\rm H}\bd{\chi}^\prime_{\max} \right)\label{e:LLRH}\ea
Substituting (\ref{e:H1}) and (\ref{e:H2}) into (\ref{e:LLRH}) yields
 \be L({\cal H}) = \left(\mbox{ABS}~(\hat{\bd{\chi}}) - \mbox{ABS} (\widehat{\bd{\chi}^\prime}) \right) {4 \alpha^2}/{N_0}\ee
where $\mbox{ABS}(\bd{a}) = \sum |a_i|$. Consequently, the
probability of hypotheses ${\cal H}_1$, ${\cal H}_2$ can be obtained from
$L(H)$ \be \Prob({\cal H}_k | \bd{y}) = {1}/{\left(1 + \exp(\mp L(H))\right)},
~k=1,2. \ee 
For each winning hypothesis  one can apply the lattice
detection algorithm of Section~{\ref{sec:detector}}---simply treat the information bearing vector
$\bd{\chi}$ as from a lattice with generator matrix $\bd{B}$,
i.e., $\bd{\chi} {=} \bd{B}\bd{u}$; e.g., the equivalent model for
detecting lattice point $\bd{\chi}$ is $
\hat{\bd{\chi}} {=} \bd{B}\bd{u} {+} \tilde{\bd{n}}^1 $, where
$\hat{\bd{\chi}}$ is the output of matched filtering of hypothesis ${\cal H}_1$. Since $\bd{\chi}$ is from a  $D_4$ lattice, its
generator matrix is $\bd{B}$ of
(\ref{e:D4}). APPs are obtained per Section~{\ref{sec:detector}}.

\subsection{Iterative Receiver for  Coordinate Interleaving in Fast Fading}
\label{itercoord}
Coordinate interleaving, along with the outer iteration loop in Fig.~\ref{icmmse}, is now considered; the real and imaginary parts of all complex symbols in a frame are collectively scrambled before transmission \cite{ion:int}.
$\bd{Y} = \{ \bd{y}_1, \bd{y}_2,\cdots,\bd{y}_N
\}$ denotes  a frame spanning $N$ MIMO channel uses at the MIMO channel output (before deinterleaving).
Note that the structure of the superorthogonal lattice code is
removed during transmission, and has to be recovered before detection. The applicable receive equation is \rf{e:realtxmodel}
rather than \rf{e:superothogonal}; the iterative IC-MMSE attempts to iteratively remove the cross-antenna interference, i.e.\ to undo the channel $\bd{H}$ on a per MIMO channel use basis. During the first iteration, the soft feedback
from the detector/decoder is null. The output of IC-MMSE is
always deinterleaved, thus restoring the superorthogonal structure and yielding the soft-output $\hat{\bd{X}} = \{
\hat{\bd{x}}_1,\hat{\bd{x}}_2,\cdots,\hat{\bd{x}}_N \}$ with
\be
\label{e:fast} \hat{\bd{x}}_t = \bd{\Gamma}\bd{\chi}_{\oplus;t} +
\tilde{\bd{n}}_t. \ee 
Since the information-bearing vector
$\bd{\chi}_{\oplus;t}$ is a direct sum of two $D_4$ lattices, and
the effective channel gain matrix $\bd{\Gamma}$ is unitary, the equalization approach in Section~\ref{quasistaticrx} applies to eq.\ \rf{e:fast}. $\Prob({\cal H}_k | \hat{\bd{x}}_t), ~ k = 1,2$, are associated with the following transmission models upon removing
$\bd{\Gamma}_1$, $\bd{\Gamma}_2$ respectively:  
\be {\cal H}_1:  \tilde{\bd{\chi}}_t =  \bd{B} \bd{u}_t
+ {\tilde{ \bd{n}}}^1_t \label{e:fastMIMO1} \ee
\be {\cal H}_2:   \tilde{\bd{\chi}}^\prime_t  =  \bd{B}\bd{u}^\prime_t + {\tilde{
\bd{n}}}^2_t \label{e:fastMIMO2} \ee 
where $\tilde{\bd{\chi}}_t =
\frac{1}{2}\bd{\Gamma}^T_1 \hat{\bd{x}}_t$, $\tilde{\bd{\chi}}^\prime_t
= \frac{1}{2}\bd{\Gamma}^T_2 \hat{\bd{x}}_t$, ${\tilde{
\bd{n}}}^1_t = \frac{1}{2}\bd{\Gamma}^T_1 \tilde{\bd{n}}_t $ and
${\tilde{ \bd{n}}}^2_t = \frac{1}{2}\bd{\Gamma}^T_2
\tilde{\bd{n}}_t$. The generator matrix $\bd{B}$ is given in
(\ref{e:D4}). For each hypothesis, the lattice decoding algorithm
can be applied to compute the extrinsic APPs $P(u;{\rm O})$ and
$P(c;{\rm O})$.

Inner-loop iterative decoding between SISO and BP, as shown
in Fig.~{\ref{icmmse}}, can  further improve the overall
performance, especially in the presence of FEC coding, when decoding follows detection. 
Herein,  only an uncoded system is considered in order to illustrate the concept. 
Even then one can perform inner loop iterations between $P^{\rm BP}(c;{\rm O})$ from the BP module and $P(u;{\rm I})$ from the SISO block; more benefit is derived when a decoder is part of the inner-loop.

\section{Simulation and Discussions}
\begin{figure}[bt]
\hskip-.1in\vskip-.1in\epsfig{file = 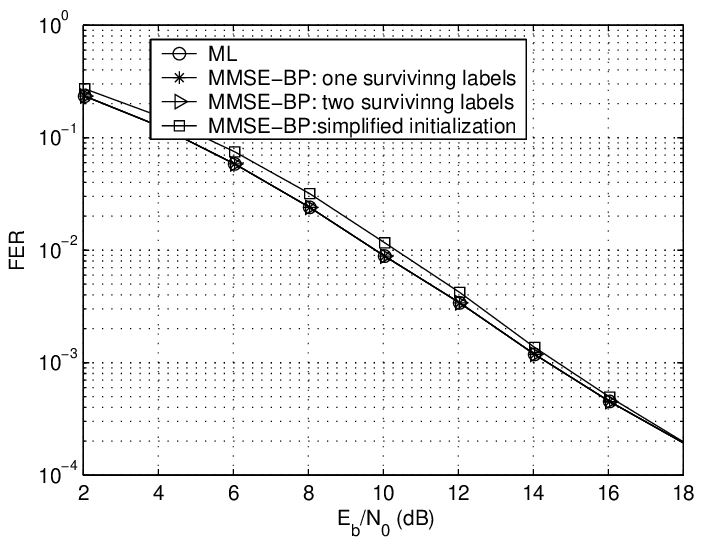, width=0.5\textwidth} \vskip-.14in\caption{Super-orthogonal space-time lattice code, MMSE filtering followed
by BP. ML and BP with one and two surviving labels are identical.
Curve with square markers illustrates effect of simplified
initialization.} \label{f:Quasi}
\end{figure}

\begin{figure}[hbt]\hskip-.18in \vskip-.14in
\epsfig{file=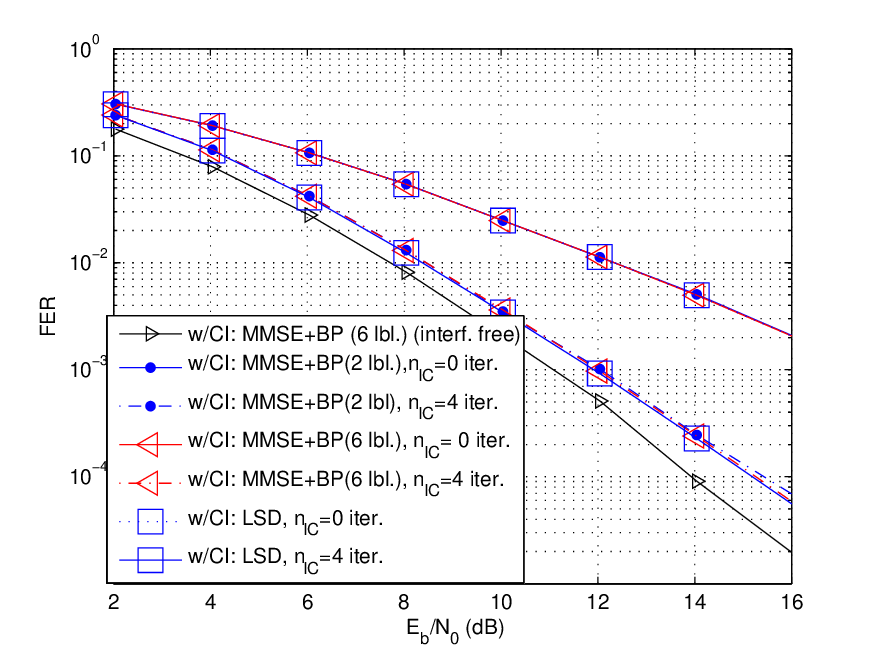, width=0.52\textwidth} \vskip-.16in\caption{Iterative decoding based on MMSE-IC followed by BP for
super-orthogonal space-time lattice code with coordinate interleaver; $n_{\rm IC}$ denotes the number of iterations in the outer loop (feedback to MMSE-IC).} \label{f:fast} 
\end{figure}

The decoding of a 32-point super-orthogonal space-time lattice constellation 
 $\mathcal{C}\in \mathbb{R}$ (Example~\ref{ex2}) using the above
algorithm is illustrated in both iterative and non-iterative scenarios. Each half of $\mathcal{C}$ 
belongs to a $D_4$ lattice, implicitly defining a shaping region; only six
of the twelve $\bd{L}(\Lambda)$ labels  listed in Example~\ref{ex3}
(see Table \ref{tab:cardin}) are needed to cover the lattice points in the
shaping region. The algorithm's efficiency was separately tested by retaining (post BP)  only
the most likely label (or two labels);  others receive zero probabilities
(re-normalization performed after setting to zero the 
probabilities of  discarded labels).
Bit 5-tuples are mapped to one of the
thirty-two codewords, 
then transmitted over two channels
uses (2.5 bits/channel use).
A (depth-eight, block) coordinate interleaver  can  scramble the coordinates of
many space-time codewords before transmission. If $b$ bits are sent per channel use, and $E_s$ is the symbol energy then
$E_b/N_0[dB]  =   E_s/N_0[dB] + 10*\log_{10}\left(({N_r}/{b})\right)$ \cite{dis:zhu}. 
The channel is constant over $T = 2$ uses. A data packet has 500 super-orthogonal
codewords; a plotted point in Figs.~{\ref{f:Quasi}},~{\ref{f:fast}} averages 2000  
packets.
Fig.~\ref{f:Quasi} shows the frame error rate (FER)\footnote{Here,
frame refers to one super-orthogonal space-time codeword.} vs.\
$E_b/N_0$ for both ML and MMSE-BP algorithm when  coordinate
interleaving  is absent ($\pi_{\rm CI}$ is the identical permutation). The ML algorithm searches all possible valid
codewords then selects the one that maximizes the likelihood; ML's optimal performance serves as
 benchmark. The MMSE-BP algorithm 
runs one\footnote{Depending on the structure of the underlying lattice, additional BP iteration(s) may be warranted; see also \cite[ Example 7.3, $\rm L^1_{256}$ vs.\ $\rm L^2_{256}$]{sad:low}.} full iteration on the TG (complete, two-way message passing between check and variable nodes), and collects the
probabilities of label coordinates. The options of retaining one surviving
label and two surviving labels---i.e., the most likely one(s)---are examined. 
The MMSE-BP algorithm achieves 
ML performance, regardless of the number of surviving labels.
When simplified initialization is used for BP on the lattice TG (reduced complexity) and two surviving labels are retained, a 0.5dB performance loss is observed relative to ML
at low SNR. As SNR increases, the loss is gradually
reduced and the algorithm using simplified initialization matches the ML performance
asymptotically.

Fig.~\ref{f:fast} illustrates the iterative
scenario with coordinate interleaving, per Fig.~{\ref{icmmse}}. Two inner iterations are run
between the SISO module and the BP module; one full iteration is run on
the lattice TG inside the BP block.  Ideal, interference-free MMSE-BP is shown as performance benchmark---the feedback vector $\bd{x}_e$  identically replicates  the transmitted vector in the first  IC pass.
When  feedback from
outer decoder to inner decoder is absent,  performance with six surviving
labels is slightly better than that with two surviving labels, but
it is about 3 dB away from interference-free MMSE-BP  at 
$\rm FER=0.005$. After four iterations between inner decoder and outer
decoder, MMSE-BP performance is improved; the subtle difference between two
 labels and six surviving labels  suggests
that practical simplifications are possible as lattice dimensionality increases. 
LSD \cite{Hochwald:LSD}, shown for performance comparison, essentially tracks 
MMSE-BP;   a slight difference from \cite{Hochwald:LSD} is that the sphere was centered on $\hat{\bd{x}}$ (see \rf{xhati}) instead of the unconstrained ML estimate $\hat{\bd{s}}$ of \cite[eq.\ (15)]{Hochwald:LSD}.

As stated in Remark~\ref{remark2}, the extrinsic information associated with coordinates of lattice points (rather than with labels)---as generated at the BP module output(s)---allows one to naturally expand the iterative  decoding structure in Fig.~{\ref{icmmse}} so as to accommodate, in an iterative receiver, a separate FEC code that might be used with  any (MIMO) lattice constellation. 
This is achievable by employing the extrinsic soft-information supplied at the outputs of the BP module in manners similar, e.g., to hybrid concatenations of SISO modules \cite{BenedettoDMP:97}; puncturing the FEC code or adjusting its rate, or the constellation order, fit naturally with the soft-output lattice detector.

\subsection{Relation with State-of-the-Art\label{sota}}
For comparing the proposed algorithm and the state-of-the-art in sphere decoding, some common ground with enough flexibility for comparisons needs to be established. The sequel (loosely) denotes by $n$ the size of an observation vector (e.g.\ $\bd{y}$ in \rf{rxeq}), and the lattice dimensionality by $m$---not discerning real from complex dimensions, since $O(\kappa g){=}O(g)$ if $\kappa{\neq}0$.

The IC-MMSE soft-equalizer is not central {\em per se} to the algorithm  developed in Sec.\ \ref{sec:detector}, as it only aims to 
avoid the distorting effects of the channel matrix on the lattice generating matrix (see discussion leading to Sec.\ \protect\ref{sec:lattdetmodel});
as discussed (ibid.) this is not a limitation in principle of the proposed search philosophy, since the  lattice structure could be inferred in real time (as was the case in \cite{Hochwald:LSD,vit:uni2,stu:sof}).
Such additional complexity cost at the receiver---in response to the fact that the underlying lattice does change with every use of the channel---would have a  complexity comparable to the $O(m n^2)$ computations of computing the {\em unconstrained} LS estimate or  similar methods (see below), i.e.\ not exponential in $m$. Thus, the IC-MMSE equalizer (if used) will not dominate algorithmic complexity.

\subsubsection{Comparison Space\label{comparspace}}
Two related approaches exist in the literature at the manuscript's time of submission: the mature sphere decoding algorithm (Fincke and Pohst), and a reported use of a lattice model for a high-dimensional LDPC block code in order to perform message passing on its lattice TG \cite{sad:low}; the latter was comparatively assessed in Sec.\ \ref{sec:intro}.

Sphere decoding is, in essence, an efficient algorithm for solving an integer LS problem like $\min_{\bd{s}\in \mathcal{Z}^m \cap\mathcal{M}} \|\bd{y} - \bd{H} \bd{s} \|^2$ (real case), 
i.e.\ finding  the LS solution  in the countable (constellation) set $\cal M$ to the problem
\be 
\bd{y}=\bd{\rm H} \bd{s} + \bd{n} \label{rxeq}
\ee
When speaking of \rf{rxeq} as an integer LS problem $\bd{y} {\in} \mathbb{C}^{n\times 1}$,  $ \bd{\rm H} {\in} \mathbb{C}^{n\times m}$, $\bd{s} {\in} {\cal M} {\subset} \mathbb{Z}^m  {+} j \mathbb{Z}^m $, and $\cal M$ is a constellation in $m$ (complex\footnote{If using a real equivalent model,\,$\bd{\rm H} {\in} \mathbb{R}^{2n{\times} 2m}$, $\bd{y} {\in} \mathbb{R}^{2 n{\times} 1}$, $\bd{s} {\in} {\cal M} {\subset} \mathbb{R}^{2 m} $.}) dimensions.
Often, ${\cal M} \subset \mathbb{Z}^m$ is  a cubic or skewed integer lattice, hence the name, but in the most general receive model \rf{rxeq} the countable constellation $\mathcal{M}$ can be some other lattice, or no lattice at all---in which case `rounding,' (or `slicing,' discussed below) would occur to the closest $\mathcal{M}$ point, instead of integer.
As well-known, zero forcing (ZF) equalization approaches demodulation by first attempting to reverse the channel effect via the channel pseudoinverse matrix, 
$\bd{\rm H}^{+} \doteq \piM{H}$, although this generally results in noise enhancement; the {\em unconstrained} LS solution 
$\hat{\bd{s}}_{\rm ml}\df \bd{\rm H}^{+}\bd{y}$ need not  belong to the finite modulator constellation $\cal M $, and would have to be reduced modulo $\cal M$ by finding the $\mathcal{M}$ point closest to $\hat{\bd{s}}_{\rm ml}$. 
When the lattice $\mathcal{M}$ verifies $\bd{s} \in {\cal M} \subset \mathbb{Z}^{m} $, such  reduction simply  amounts to replacing each coordinate of the unconstrained ZF solution by the closest valid integer, which is also known as the Babai estimate. Unfortunately, for the modulo-$\mathcal{M}$ problem, slicing the unconstrained LS ML solution is  generally not optimal w.r.t.\ $\mathcal{M}$ in the ML  sense. Nevertheless, ML optimality {\em is} preserved by  reduction modulo $\mathcal{M}$ (rounding, or `slicing'), to  a {\em limited extent}, in \textit{some} cases---i.e., when $\bd{H}$ of \rf{rxeq} has orthogonal columns.
ZF may be improved upon by decision feedback or cancellation and ordering (or VBLAST, see discussion on sphere decoding literature in Sec.\ \ref{sec:intro}).

\newtheorem{prop2}{Proposition}
\begin{prop2}\label{prop2}[ML optimality of $\mathcal{M}$-constrained solution to \rf{rxeq}]
If $\bd{\rm H}$ in \rf{rxeq} 
has orthogonal columns 
then finding the closest $\mathcal{M}$ point to the unconstrained LS ML  solution to \rf{rxeq} preserves optimality in the {\em constrained} ML sense. 
\end{prop2}
\begin{proof} Appendix~\ref{prfprop2} sketches, for quick reference, a self-contained proof of this otherwise known result, stressing that {\em $\mathcal{M}$ need not be an integer lattice, or even a lattice at all}.\end{proof}

Finally, the search sphere center may be placed (albeit not necessarily \cite{vit:uni1})  at the unconstrained LS solution. This is the case with  LSD \cite{Hochwald:LSD},
and we proceed from this perspective---in order to facilitate certain points of comparison with the LSD algorithm, which is the closest counterpart when concerned with MIMO lattice constellations, and with incorporating soft-information in the lattice search algorithm. One relevant aspect concerns the initialization of the label probabilities (Sec.\ \ref{sec:initlblprob}). Soft-outputs were also pursued by Boutros {\em et al.}\ \cite{bou:sof} and by Studer {\em et al.}\ \cite{stu:sof}, who compute bit log-likelihood ratios by refining a tree-traversal strategy (cf.\ {\em op.\ cit.}\ and references)---but do not accommodate {\em a priori} information; \cite{stu:sof} used  Schnorr-Euchner's version of Pohst's algorithm \cite{agr:clo} with radius reduction (whereas LSD keeps the radius constant).
\subsubsection{Differentiation and Advantages of Proposed Algorithm\label{sec:adv}}
The new idea 
is to approach a soft-output search within a finite set of points carved from a lattice as a 3-pronged strategy:
\begin{enumerate}
\item replace the (more or less) {\em ad hoc} sphere method of containing the search space by a mechanism that naturally, inherently, and automatically leverages on the intimate structure of the lattice (vis-\`a-vis the label cosets), 
\item avoid inadvertently discarding the closest lattice point, seek the maximum {\em a posteriori} (MAP) point `around' it;
\item parse the intrinsic mechanisms that can generate extrinsic information on lattice points or coordinates into two tiers: 
\begin{dinglist}{43}\item one that quantifies the mutual information between lattice points---or their coordinates---on one hand, and the labels on another hand (the Markov process of Sec.\ \ref{pointAPP}, or SISO module in Fig.\ \ref{icmmse}); and \item another that reflects the source of redundancy, i.e.\ the underlying Abelian block label code (BP {\em per se}). \end{dinglist}
\end{enumerate}

\noindent In one aspect, the burden of efficiently confining the search volume within the $m$-dimensional lattice space---e.g., by finding and managing a search radius, or a parallelogram \cite{kan:imp}, or a candidate list (in LSD), in order to preclude an exponential increase in complexity---is naturally replaced in the proposed algorithm by monitoring one or more labeled cosets (the labels being from the lattice label code $\bd{L}(\Lambda)$). Modifying the search radius in the classic  sphere decoding algorithm simply corresponds to modifying the number of monitored (surviving) labels; this is, of course, the case irrespective of whether the lattice generating matrix changes with the channel use, or not.

\newtheorem{rem3}[rem1]{Remark}
\begin{rem3}
\label{remark3}
In the view of \cite{has:sph}, sphere decoding {\em per se} need not decisively address  the initial determination of a sphere radius, which is left to other mechanisms---e.g., use of the lattice covering radius (NP hard), or of a value arising from a noise model. \cite[Sec.\ IV.A]{has:sph} does propose to iteratively discover a good radius based on a noise confidence interval---should an initial setting yield no lattice point; this hints to {\em some} correlation between radius and the lattice geometry itself. LSD, too, advocates the estimation of a noise mean power---albeit, in addition to a radius component that models  the lumped effect of both the channel matrix and (indirectly) the lattice generating matrix \cite[eq.\ (28)]{Hochwald:LSD}
\hfill $\square$
\end{rem3}

It is now apparent that the method of Sec.\ \ref{sec:detector} naturally links the size of the reduced search space  to lattice geometry via the inherent structure of the labeled cosets (rather than a sphere). Ultimately, confining the search space is efficient only to the extent that it does not inadvertently eliminate the closest lattice point, and can keep enough of its neighbors to include nearby points that may maximize the MAP probability. (For that same reason, LSD does not reduce the search radius as new lattice points are found.)

Understanding this aspect requires another look at the physical meaning of the unconstrained LS solution computed w.r.t.\ $\Lambda$---when viewed {\em from the perspective of the projections} $P_{W_i}(\Lambda)$, 
$i{=}1, \ldots, m$. 
By virtue of Proposition \ref{prop2}, although $P_{\mathcal{W}}\Lambda\doteq \bd{W}^{\rm T}\Lambda$ is not a sub-lattice, the LS ML solution constrained to 
$\mathcal{M}\doteq P_{\mathcal{W}}\Lambda$ is, nonetheless, the closest point from $P_{\mathcal{W}}\Lambda$ to the orthogonal projection---onto 
$\mathcal{W}$---of $\bd{y}$\footnote{If ${\cal M} {\subset} \mathbb{Z}^m $ in \rf{rxeq} then $\bd{\rm H}$ plays the role of a lattice generating matrix, and may include the effect of some channel matrix, lumped with the true lattice generating matrix; either way, $\bd{\rm H} \bd{s}\doteq \breve{\bd{\lambda}}$ can be viewed as a point in some skewed lattice $\breve{\Lambda}$. Since $\bd{y}=\bd{\rm W} {\bd{\rm W}^{\rm H} \breve{\bd{\lambda}}} + \bd{n} = \bd{W} P_{\mathcal{W}}\breve{\bd{\lambda}}+\bd{n}$, with $\bd{W}$ orthogonal, the LS ML solution constrained to $P_{\mathcal{W}}\breve{\Lambda}$ is simply the closest point from $P_{\mathcal{W}}\breve{\Lambda}$ to the projection of $\bd{y}$ onto $\mathcal{W}$. \protect\label{zfinvarproj}}; $\bd{y}$ could be the unconstrained LS solution w.r.t.\ $\Lambda$, or $\hat{\bd{x}}$ of Sec.\ \ref{sec:equalizer}. Since $\bd{W}$ is orthogonal, the projections $P_{W_i}(\Lambda)$ can be obtained dimension-wise, with complexity $O(m)$. More importantly, the coordinates of the {\em constrained} ML point in $P_{\mathcal{W}}\Lambda$ are, thereby, obtained---as in Sec.\ \ref{sec:projdmn}---simply as the closest points in $P_{W_i}(\Lambda)$ to the respective projections on $W_i$ of some unconstrained  ML statistic (e.g., the unconstrained LS solution w.r.t.\ $\Lambda$, or $\hat{\bd{x}}$).

\newtheorem{ex1}[ex2]{Example}
\begin{ex1}
\label{ex1}
Computing $\Prob(\bd{l})$ of \rf{prlbl1}, \rf{prlbl2} depends on the (squared) distances  in \rf{mincoorddist} respectively \rf{mincoorddistS}. When \rf{prlbl1}  is used, a squared distance as in \rf{mincoorddist} must be computed as many times as $\sum_{\bd{l}}(|\Lambda(\bd{l})| m)= m \sum_{\bd{l}}|\Lambda(\bd{l})| = m |\Lambda|$; in the latter case a squared distance is computed $\sum_{\bd{l}}\sum_{i=1}^m|{\cal S}_i (\bd{l})|$ times (see Remark \ref{remark1}). For the superorthogonal constellation at hand---taking into account that one of the two direct-sum {\em lattice} terms $D_4$ is selected prior to label probability computations---$|\Lambda|{=}16$, $m{=}4$, and $|{\cal S}_i (\bd{l})|$ are listed in Table~\ref{tab:cardin} (only six labels are needed for the shaping region $\cal R$).
\begin{table}[hb]\caption{$|{\cal S}_i (\bd{l})|$ vs.\ $\bd{l}\in \bd{L}(\mathcal{C}(\Lambda,\bd{u}_0,\mathcal{R}))$\protect\label{tab:cardin}}
\centering\begin{tabular} {@{\vline}c@{\vline}c@{\vline}c@{\vline}c@{\vline}c@{\vline}c@{\vline}} \hline 
\multicolumn{1}{@{\vline}p{.4in}@{\vline}}{\centering $\bd{l}$} 
& \multicolumn{1}{@{}c@{\vline}}{\centering $\ |{\cal S}_1 (\bd{l})|\ $} & \multicolumn{1}{@{}c@{\vline}}{\centering $\ |{\cal S}_2 (\bd{l})|\ $} & \multicolumn{1}{@{}c@{\vline}}{\centering $\ |{\cal S}_3 (\bd{l})|\ $} & \multicolumn{1}{@{}c@{\vline}}{\centering $\ |{\cal S}_4 (\bd{l})|\ $} & \multicolumn{1}{@{}c@{\vline}}{\centering $\ |\Lambda (\bd{l})|\ $}\\ [5pt]\hline\hline
 0440 & 3 & 2 & 1 & 1 & 3\\ [3pt] \hline
 0411 & 3 & 2 & 1 & 1 & 3\\ [3pt] \hline
 0251 & 3 & 2 & 1 & 1 & 3\\ [3pt] \hline 
 0220 & 3 & 2 & 1 & 1 & 3\\ [3pt] \hline
 0031 & 1 & 1 & 1 & 1 & 2\\ [3pt] \hline
 0000 & 1 & 1 & 1 & 1 & 2\\ [3pt] \hline
\end{tabular}\end{table}
Use of \rf{prlbl2} can reduce complexity---to an extent that generally depends on the actual lattice---by a factor ranging, in this case, from  $\frac{64}{36}$ (when all labels are used, i.e.\ no attempt is made to reduce the search space)  to $\frac{64}{7}$ (one surviving label is preserved); e.g., if two surviving labels are preserved then the complexity reduction is by a factor between $\frac{64}{14}$ and $\frac{64}{8}$. 
\hfill $\square$
\end{ex1}

The importance of this aspect is appreciated when initializing label probabilities: it insures that the largest likelihood is preserved among the projections $P_{W_i}(\Lambda)$---even when the max-log approximation is used (for coset probabilities). This guarantees that  the coset containing the ML lattice point is promoted as the most likely coset---when iterations start. In restricting the search to the points labeled with the most likely label(s), the algorithm will not inadvertently discard the closest lattice point. This remains true if more than one label is retained---given  that the closest lattice point may not be the one that maximizes {\em a posteriori} extrinsic information.\footnote{Even if the likelihood order for the remaining points (other than the ML one) is spoiled by the constrained LS solution, the redundancy present in the lattice and/or FEC will iteratively restore the {\em a posteriori} order, as long as the algorithm proceeds with the correct search scope.}
\newtheorem{prop3}[prop2]{Proposition}
\begin{prop3}\label{prop3}
No information about the closest lattice point is lost in restricting the search to the lattice points associated with the most likely labels(s). The algorithm in Sec.\ \ref{sec:detector} will not inadvertently discard the closest lattice point, and the additional points associated with the most likely label(s) provide for the possibility that the closest lattice point may not be the one that maximizes {\em a posteriori} extrinsic information.
\end{prop3}

At last, the iterative discovery of the MAP solution is allocated between two functionalities: one that quantifies the mutual information between labels and the {\em labeled} points, and one that iterates on label probabilities (BP on the TG).

In conclusion, by replacing the sphere---in the classic sphere decoding or LSD algorithms---with one or more label(s), or coset(s), the management of the search space is greatly simplified, without losing information about the closest lattice point, and with well-behaved complexity (see next subsection). 

\subsection{Concerning Algorithmic Complexity\label{algcompl}}
Apart from label probability initialization---which is related to finding an intermediate solution to the constrained problem w.r.t.\ $P_{\mathcal{W}}\Lambda$, and has complexity $O(m)$, see above---complexity is influenced by SISO APP and BP on the label TG.


The generic SISO MAP algorithm is  known to have a complexity linear in the state complexity of the trellis \cite{mon:des}. 
State complexity $s$ is a fundamental descriptive characteristic of a (block) code, comparable to its length, rate, and minimum distance (cf.\ \cite{mud:min,laf:low} and references). Notably, when normalized to the block length---and referred to as relative trellis complexity $\varsigma$---it coincides with all other measures of complexity (branch- or edge-complexity, total number of merges, etc.) asymptotically as the length tends to $\infty$ \cite[Sec.\ VI, p.\ 1952]{laf:low};
\footnote{Thereby, although the discussion continues below in terms of $s$, a parallel limiting behavior argument can be made w.r.t.\ {\em all} measures of {\em relative}, or {\em normalized}, complexity via $\varsigma$---perhaps more meaningfully, because normalization relates  the complexity to some physical symbol duration.}
 it is the asymptotic behavior, after all, that matters in the limit as $m\rightarrow \infty$.

State complexity is upper bounded---for both linear \cite[Sec.\ IV]{mud:min}, \cite[eq.\ (1)]{laf:low} and nonlinear\footnote{\hspace{0pt}The minimization over all one-to-one trellises \protect\cite[eq.\ (2)]{laf:asy} is superfluous if the trellis on which decoding is to be performed is given.} trellises \cite[eq.\ (2)]{laf:asy}, \cite{laf:low}---by $\max_{i\in \mathcal{I}}\left(\log_q \left|S_i\right|\right)=\max_{i\in \mathcal{I}}\left(\log \left|S_i\right|\log_q 2\right)\leq \max_{i\in \mathcal{I}}\log |S_i|$, 
where $q\geq 2$ denotes  edge label alphabet size and $S_i$ is simply the set of states at time (trellis section) $i$. 

Because the number of states in the SISO module equals the number of labels (Sec.\ \ref{pointAPP}), complexity will be eventually determined by the logarithm of the number of labels $\log(|\bd{L}(\Lambda)|)$. Per \cite[Sec.\ V-B.1]{ban:tan}, $|\bd{L}(\Lambda)|=|\bd{G}|/|\bd{L}(\Lambda)^\ast|\leq |\bd{G}|$, with $|\bd{G}|=\prod_{i=1}^m g_i$; take $|\bd{G}|$ as a measure of complexity, as done in fact in \cite{ban:tan}, too. Using \cite[Prop.\ 6, Corollaries 2, 3]{ban:tan}, and \cite{ban:tre},  the limiting behavior of  decoding complexity as $m\rightarrow \infty$ is captured in Table \ref{tab:complxvslatt}  for several important lattices.

Thereby, there exist lattices for which complexity is not exponential in $m$; contrast this with the mean algorithmic complexity of classic sphere decoding, shown \cite[Sec.\ III.B.1]{has:sph} to be exponential in lattice dimensionality---in the sense of the mean number of visited points (fixed lattice, arbitrary noise). 
\begin{table}[hb]\caption{Asymptotic complexity in lattice  dimensionality $m$ for checkerboard, Barnes-Wall and root lattices\protect\label{tab:complxvslatt}}
\centering\begin{tabular} {@{\vline}p{.9in}@{\vline}p{2.2in}@{\vline}p{.3in}@{\vline}c@{\vline}c@{\vline}c@{\vline}} \hline 
\multicolumn{1}{@{\vline}p{.9in}@{\vline}}{\centering lattice $\Lambda$} 
& \multicolumn{1}{@{}p{2.2in}@{\vline}}{\centering $\log(|\bd{G}(\Lambda)|)=\log\left(|\bd{L}(\Lambda)| |\bd{L}(\Lambda)^\ast|\right) \in \ldots $} & \multicolumn{1}{@{}p{.3in}@{\vline}}{\centering \hspace{0pt}Reference } & \multicolumn{1}{@{}c@{\vline}}{\centering } & \multicolumn{1}{@{}c@{\vline}}{\centering  } & \multicolumn{1}{@{}c@{\vline}}{\centering  }\\ [5pt]\hline\hline
 ${\rm D}_m$ &  $O(m)$ & \cite{ban:tan}  &   &   &  \\ [3pt] \hline
 ${\rm BW}_m$, $m\doteq 2^{m_0}$, $m_0$ even & $O\left( \log m^{{m}/{2}}\right)=O(m \log m)$  & \cite{ban:tan}  &   &   &  \\ [3pt] \hline
 ${\rm BW}_m$, $m\doteq 2^{m_0}$, $m_0$ odd  & $O\left( \log (2^{-m/2} m^{{m}/{2}})\right)=O(m \log m)$ & \cite{ban:tan}  &   &   &  \\ [3pt] \hline 
 ${\rm A}_m$ &  $(m \log m)/2 {+}O(m) + (m \log m)/2 {+}O(m) = O(m \log m)$  & \cite{ban:dis}  &   &   &  \\ [3pt] \hline
\end{tabular}\end{table}

A simple observation vis-\`a-vis the lattice coding gain $\gamma(\Lambda)$ \cite{con:sph}, \cite[eq.\ (9)]{ban:tan} and Hermite's number theoretic constant $\gamma_m\doteq 4 \delta_m^{2/m}$ \cite[p.\ 20]{con:sph}---with $\delta_m$ being the {\em densest} lattice packing in $\mathbb{R}^m$---allows one to extend the above conclusion to general lattices.  The gain of an arbitrary lattice is $\gamma(\Lambda)\doteq \lambda^2(\Lambda) [V(\Lambda)]^{-2/m}$, where $V(\Lambda)$ denotes the volume of the fundamental (Voronoi) region\footnote{Caution  as to the definition of $V(\Lambda)$; in \cite[p.\ 4]{con:sph} it is the {\em square root} of $\det \Lambda$, as opposed to $\det \Lambda$ in \cite[eq.\ (9)]{ban:tan}. This is inconsequential, as the respective definitions of $\det \Lambda$ revert the discrepancy, and in both cases $V(\Lambda)$ equals the determinant of the generating matrix---if latter is square.}, and $\lambda(\Lambda)$ is the shortest nonzero vector in $\Lambda$. Per \cite[Corollary 3]{ban:tan}\footnote{\label{eqinmanylatt}Equality achieved for many known lattices and optimal coordinates\cite{ban:tan}.}, 
\be \textstyle |\bd{G}|\geq \left\lceil \gamma (\Lambda)^{1/2} \gamma (\Lambda^\ast)^{1/2}\right\rceil^m \label{Gbound}\ee
\noindent It can be shown that $\gamma(\Lambda)= 4 \delta^{2/m}$ \cite[p.\ 73 \& eq.\ (20)]{con:sph}, where $\delta$ is $\Lambda$'s  center density; but $\gamma$  {\em lower}-bounds  
$\gamma_m$---as $\delta_m$ is the center density of the {\em densest} lattice---which in turn verifies $\frac{\gamma_m}{m} \lesssim \frac{1.744}{2 \pi e}\doteq C$ for large $m$ \cite[eq.\ (48)]{con:sph}. Applying this inequality to both $\gamma(\Lambda)$, $\gamma(\Lambda^\ast)$ yields\footnote{Rigorously speaking, we assume that $\Lambda$ is appropriately represented so as to achieve equality in \rf{Gbound}, or a tight enough approximation; e.g.,  $|\bd{G}|\in O\left(\left\lceil \gamma (\Lambda)^{1/2} \gamma (\Lambda^\ast)^{1/2}\right\rceil^m\right)$ would suffice to assess limiting behavior. This is possible at least for some `good' known lattices, see footnote \protect\ref{eqinmanylatt}.} $|\bd{G}|{\lesssim} \left(m C\right)^{m}$ and
\be \log |\bd{G}(\Lambda)|\in O(m \log m + m \log C)=O(m\log m)\label{gencmplxty}\ee
Complexity is at most log-linear in  lattice dimensionality $m$---and better than cubic ($O(m^3)$, {\em cf.}\  \cite{has:sph,Hochwald:LSD}), for arbitrary\ $\Lambda$.

Lastly, the complexity of BP on the TG was characterized in \cite{ban:tan}---measured by $|\bd{G}(\Lambda)|$, whose limiting behavior was discussed above. Alternatively, it is well known that a lattice can be decoded on its trellis---which is well-defined, finite, and based on its quotient group $\Lambda/\Lambda^\prime$ \cite{for:den}. The same discussion on state complexity applies, with the slight difference that now complexity would be measured by $|\bd{L}(\Lambda)|<|\bd{G}(\Lambda)|$. Each distinct path through the trellis corresponds to a coset, and  in the limit with $m\rightarrow\infty$ the complexity of BP on the TG cannot be worse than log-linear, per above discussion.

Thereby, the aggregate complexity of the proposed algorithm is  at most log-linear in lattice dimensionality.

\section{Conclusion}
A soft-output MAP point search in lattices was introduced, via a form of  BP on a lattice. Due to the coding gain associated with a lattice, structural relations exist between certain lattice points, which can be associated via an equivalence relation for detection purposes. This leads to a soft-output detection algorithm, which can generate both total and extrinsic  {\em a
posteriori} probability  at the detector's output. The new idea 
is to approach a soft-output search within a finite set of points carved from a lattice as a 3-pronged strategy, as discussed in Sec.\ \ref{sec:adv}. The backtrack artifact of classic sphere decoding is absent, and complexity is at most log-linear in the lattice dimensionality.

{\appendices
\section{Finding extrinsic {\em a posteriori} probabilities post BP}
\label{app1} Herein,  the expressions for extrinsic a posteriori
probabilities \rf{pcotg}, \rf{puotg}, at the BP  detector's output, are derived; the extrinsic
probabilities are needed in iterative receivers.
Here, the goal of detection is to provide soft information about valid channel alphabet symbols, i.e.\ real coordinates of the complex symbols from the modulation constellations used on various transmit antennas; this information about coordinates can be used to revert the effect of a coordinate interleaver, or can be forwarded directly to a soft decoder for some coded modulation encoder. Alternatively, it can be used for soft or hard demodulation, e.g.\ in the case of bit interleaved coded modulation, or with plain uncoded transmission.

When a lattice is represented by a TG, it is possible to associate  a Markov process with the model for soft detection of lattice points in a natural way. This is enabled by first viewing the sequence of lattice points passed through the channel as a Markov source. Another observation is that, in general, simple detection (with or without soft information) is by itself memoryless; thereby, one should expect  the Markov process to be somehow degenerated, in order to reflect the memoryless nature of simple (non-iterative) detection. The objective of detection is to determine the a posteriori (total or extrinsic) probabilities of the output of the Markov source. In order to leverage off of known results---even in the case of plain, uncoded transmission (no FEC redundancy added by encoding)---one can view the output $\bd{c}$ of the Markov source  (a lattice point, i.e.\ a  vector of lattice coordinates) as the result of mapping with rate one (i.e.\ no additional redundancy) an identical replica of the input $\bd{u} {=} \bd{c}$; this is a degenerated Markov process where even the dependence of the future on the present is removed. The only remaining structure to be captured for the Markov source, in the case when the candidate points are from a lattice, must reflect the partitioning in labeled cosets, as discussed in Section~\ref{TG}. To this end, note that the labels themselves can be associated with states having integer values by virtue of the following convention: the state $S_{k{-}1}$ at  time $k{-}1$ is the {\em index} of the label that contains the most recent lattice point output by the Markov source, i.e.\ at time $k{-}1$; when the Markov source outputs a new point at time $k$ it transitions into  state $S_k$ equal to the integer indexing the label that contains the new point. Alternatively, with respect to the mapping  $\bd{u} {\mapsto} \bd{c}$ and omitting the time index, when $\bd{u}{=}\bd{\lambda} {\in} \Lambda$ occurs at the rate-one block input, the Markov process transitions into the state whose (integer) value  indexes the label  containing $\bd{\lambda}$. This is depicted in Fig.~\ref{markov},
where $e$ denotes an edge from starting state $s^S\!(e)$ to  ending state $s^E\!(e)$. Formally, for any edge $e$, at any time, if $\bd{u}(e){=}\bd{\lambda} {\in }\Lambda(\bd{l}_i){ \subset } \Lambda$, where $i{\in }\{1, \ldots |\bd{L}(\Lambda)|\}$ indexes one of the $|\bd{L}(\Lambda)|$ labels, then the ending state  $s^E(e){=}i$ and the Markov source outputs $\bd{c}(e){=}\bd{u}(e)$. A bijective map $\ell$ from integer states to labels $s {\map{\ell}} \bd{l}_s$ exists
such that, for any integer state $s{\in }\{1, \ldots |\bd{L}(\Lambda)|\}$, $\ell(s){\df}\bd{l}_s$ is the label associated with\,$s$.

The Markov sequence of random points selected from the lattice can be thus viewed as triggered by  state transitions triggered by $\bd{u}{=}\bd{\lambda} {\in} \Lambda$; although the realizations of $\bd{u}$ on the lattice grid are random, a state model arises as a result of partitioning the lattice in equivalence classes. That is, there exist  certain structural relations between certain points, which can be associated via an equivalence relation. The state probabilities, used in a posteriori probability calculations, are seen to be associated with the probabilities of these equivalence classes (or their labels), which  can be obtained, in turn, separately from BP on the lattice's TG, as shown next.

In general, for a Markov process generated by triggering state transitions via some input (e.g.\ a classical convolutional code), the new state depends on the current input and several previous inputs; in the case at hand the new state depends only on the current input. This illustrates the degenerated nature of the Markov process at hand,   seen thereby to be memoryless.

The memoryless nature of the Markov process is  apparent in the fact that any state can be reached in one transition from any state, and the
state probability distribution does not depend on time; it depends only on the probability distribution for $\bd{u}$, and so does the probability distribution of the output of the Markov process. The output of the Markov process does not depend on the current state, but rather on the input $\bd{u}$; the input determines both the new output and the new state---i.e.\ the output at any time does not depend on any previous state.

The state transition diagram  for the Markov process  forming the object of detection is shown in Fig.~\ref{markov}; the results in \cite{BenedettoDMP:97,bah:opt}  apply. Following \cite{BenedettoDMP:97}, the extrinsic APPs $P^{\rm BP}_k(c^j;{\rm O})$ and $P^{\rm BP}_k(u^j;{\rm O})$ during the $k$-th transition between states are 
\ba  P^{\rm BP}_k(c^j;{\rm O})&\!=\!\! & {\textstyle\sum_{e:C_k^j(e) = c^j}
A_{k-1}[s^S(e)]\prod^m_{i=1}P_k[u^i(e);{\rm I}] \nonumber }\\
 & & {\textstyle\times \prod^m_{i=1;i \neq j} P_k[c^i(e);{\rm I}]B_k[s^E(e)], \label{pco}}\ea
 \ba  P^{\rm BP}_k(u^j;{\rm O}) &\!=\!\! & \textstyle{\sum_{e:U_k^j(e) = u^j}
A_{k-1}[s^S(e)]\prod^m_{\protect\stackrel{i=1}{i \neq j}}P_k[u^i(e);{\rm I}]} \nonumber \\
& & {\textstyle \times \prod^m_{i=1} P_k[c^i(e);{\rm I}]B_k[s^E(e)], \label{puo}} \ea 
where
$A_{k-1}[s^S(e)]$ and $B_{k}[s^E(e)]$ are the probabilities of the
current state and the new state that are associated
with edge $e$. Following the well-known results and notation in \cite{bah:opt} and using the memoryless nature of the Markov process in Fig.~\ref{markov},
\ba
 A_k[s] \!\!&\!\! \df \! \!&\! \! \Prob \{S_k{=}s;  \bd{y}_1^k\} = \Prob \{S_k=s; \bd{y}_k;  \bd{y}_1^{k-1}\} \nonumber \\
 \!\!&\!\! =\! \!& \!\! \Prob \{S_k{=}s; \bd{y}_k| \bd{y}_1^{k-1}\}\Prob \{  \bd{y}_1^{k-1}\} \label{l1} \\
 \!\!\!&\! \!\!= \!\!\!& \!\! \Prob \{S_k {=}s;\! \bd{y}_k\}\Prob \{  \bd{y}_1^{k-1}\} \!\! = \! \Prob \{S_k {=} s; \bd{y}_k\} \kappa_0,
\ea
where, following \cite{bah:opt}, $\bd{y}_0^{\tau}$ denotes the observations  of the relevant Markov process, as taken at the output of a discrete memoryless channel at time instants $0,  {\ldots}, \tau$. Most importantly, the factor $\kappa_0$ does not depend on  state $s$, and is thus canceled out during the normalization step that enforces $\sum_s A_k[s]{=}1$. Due to the isomorphism between states and labels it follows that $\Prob \{S_k{=}s; \bd{y}_k\}$ is the label probability $\Prob \left\{ \ell(s)\right\}{=}\Prob \left\{ \bd{l}_s\right\}$ computed as  in \rf{e:totalAPP}. \rf{pcotg}, \rf{puotg} follow from \cite{bah:opt} and  properties of the degenerated Markov process, since
\be
B_k[s] \df \Prob \{\bd{y}_{k+1}^\tau|S_k=s\} = \Prob \{\bd{y}_{k+1}^\tau\}
\ee
does {\em not} depend on   state $s$ and behaves as a constant (canceled out during the normalization step enforcing $\sum_s B_k[s]{=}1$).

\section{}
\label{appndx}
\newtheorem{ex4}[ex2]{Example}
\begin{ex4}\label{ex4}
{\em (Linear dispersion codes)} A linear dispersion code \cite{has:hig} defines a mapping of a
complex vector $\bd{s} = [s_0,s_1,\cdots,s_{K-1}]^{\rm T} $ to a $T
\times N_t$ complex matrix $\bd{S}$ as follows: 
\be \textstyle \bd{S} =
\sum^{K-1}_{l=0} (s_l \bd{P}_l + {s_l}^{\rm H}\bd{Q}_l ) \label{complexlincomb} \ee 
where
$\{\bd{P}_l \}^{K-1}_{l=0}$, $\{\bd{Q}_l \}^{K-1}_{l=0}$ are $T \times N_t$ complex valued
matrices. The linear dispersion code can be further rearranged as 
\be \textstyle \bd{S} = \sum^{K-1}_{l=0} \left(\Re(s_l)
\bd{\tilde{P}}_l + \Im(s_l)\bd{\tilde{Q}}_l \right) \label{reallincomb} \ee 
with
$\bd{\tilde{P}}_l {=} \bd{P}_l {+} \bd{Q}_l$, $\bd{\tilde{Q}}_l {=}
i\bd{P}_l {-} i\bd{Q}_l.$ Letting $\bd{\chi} {=} \mathcal{I}(\bd{s})$ 
one can express
the linear dispersion code linearly in terms of  $\bd{\chi}$ and  a  matrix set ${\mathcal C} {\df}  \{\bd{C}_l \}^{2 K{-}1}_{l=0}{=}\!
\left\{\!\bd{\tilde{P}}_0, 
{\cdots},\!\bd{\tilde{P}}_{K{-}1}, \!\bd{\tilde{Q}}_0,
{\cdots},\!\bd{\tilde{Q}}_{K{-}1} \!\right\} $ as 
\be \textstyle \bd{S} {=} \sum^{2K-1}_{i=0} \chi_i \bd{C}_i. \ee 
If $\bd{\Gamma} \!\df\![\phi(\bd{C}^{\rm T}_0),\cdots,\phi(\bd{C}^{\rm T}_{2K-1})]$ 
  then $\bd{S}^{\rm T}$ is $\phi$-isomorphic with 
\be \textstyle \bd{x} \df \phi(\bd{S}^{\rm T}) = \sum^{2K-1}_{i=0} \chi_i \phi(\bd{C_i}^{\rm T}) = \bd{{\it \Gamma}} \bd{\chi} \label{e:chi} \ee 
\end{ex4}
Letting $\bd{\chi}$ in \rf{e:chi} be proportional to a vector of integers implies that a linear dispersion code is a lattice code with generator matrix $\bd{\Gamma}$; this is the case when $\bd{s}$ is from a particular modulation constellation, e.g.\  PAM or QAM. In general, $\bd{\chi}$ is not an integer vector, e.g.\ when the elements of $\bd{s}$ are from a PSK
constellation. However, 
if, by construction of the linear dispersion code, $\bd{s}$ is  selected to be from a lattice $\Lambda^\prime$ then the points 
$\bd{\chi}$ are carved from  $\Lambda^\prime$ via shaping region $\mathcal{R} \in \mathbb{R}^m$; i.e., 
$ \textstyle \bd{\chi} \in \bd{\Lambda}^\prime \bigcap \mathcal{R} $ where 
$\Lambda^\prime = \{\bd{B}\bd{u} : \bd{u} \in \mathbb{Z}^{m} \}$ and  $\bd{B}$ is
the generator matrix of ${\Lambda}^\prime$, while  the linear dispersion code is a lattice space-time
code with generator matrix  $\bd{\Gamma}\bd{B}$. One may find  different pairs of lattice $\Lambda^\prime$ and shaping region $\mathcal{R}$
defining the same $\bd{\chi}$s; the choice of $\Lambda^\prime$
and $\mathcal{R}$ will influence the complexity of the corresponding
decoder, as discussed in \cite{ban:tan} (unless some basis reduction approach is used to process the generator matrix).
The real transmission model becomes \be \bd{y} = \bd{H}\bd{\Gamma}\bd{B}\bd{u} +
\bd{n}, \ee
and is equivalent to using a lattice 
with generator matrix  $\bd{\Gamma}\bd{B}$.

\section{Proof of Proposition~\ref{prop2}}\label{prfprop2}
\newtheorem{prop1}{Lemma}
\begin{prop1}
\label{prop1}
A matrix  $ \bd{\rm H} {\in} \mathbb{C}^{n{\times}m}$ is left (right) diagonalizable by a unitary matrix  iff $ \bd{\rm H}$ has orthogonal columns (rows).
\end{prop1}

{\em Proof}: Let $ \rowv{x}{i}$, $ \colv{x}{i}$ be the $i$-th row, respectively column vector, of some matrix $ \bd{\rm X} $.
For the direct implication, $ \bd{\rm H} $ has orthogonal columns, thus $\bd{\rm H}^{\rm H} \bd{\rm H}$ is diagonal. By SVD, $\bd{\rm H}=\bd{\rm V} \bd{\rm \Sigma} \bd{\rm W}^{\rm H}$, where $ \bd{\rm V} \in \mathbb{C}^{n\times n}$, $ \bd{\rm W} \in \mathbb{C}^{m\times m}$ are unitary, and $ \bd{\rm \Sigma} \in \mathbb{R}^{n\times m}$ is diagonal  with non-negative diagonal entries (non-negative square roots of eigenvalues of $\bd{\rm H}\bd{\rm H}^{\rm H} $, equal in turn to $ \|\colv{h}{i}\|^2 $, $i=1 \ldots m$, per assumed orthogonality).  Thus $ \bd{\rm H}^{\rm H} \bd{\rm H}=\bd{\rm W}\bd{\rm \Sigma}^{\rm H}\bd{\rm V}^{\rm H} \bd{\rm V} \bd{\rm \Sigma} {\bd{\rm W}}^{\rm H}= \bd{\rm W} \bd{\rm D}   \bd{\rm W}^{\rm H}$, 
where $\bd{\rm D} \doteq  \bd{\rm \Sigma}^{\rm H} \bd{\rm \Sigma}= \bd{\rm \Sigma} \bd{\rm \Sigma} \in \mathbb{R}^{m\times m}$ is diagonal with diagonal elements  $d_i=\|\colv{h}{i\,}\|^2$, $i{=}1 \ldots m$, and 
$\bd{\rm H}^{\rm H} \bd{\rm H}\bd{\rm W} = \bd{\rm W} \bd{\rm D}$. But
$\forall \bd{X}$, 
\begin{eqnarray}
\bd{\rm X} 
\diag (\left[ \renewcommand{\arraystretch}{.5}\begin{array}{@{}c@{}c@{}c@{}}
\delta_1, & \cdots, & \delta_m 
\end{array} \renewcommand{\arraystretch}{1}\right])
&=&
\left[\begin{array}{@{}c@{\ }c@{\ }c@{}}
 \colv{x}{1} 
 & \ldots & \colv{x}{m}
\end{array}\right] \diag (\left[ \renewcommand{\arraystretch}{.5}\begin{array}{@{}c@{}c@{}c@{}}
\delta_1, & \cdots, & \delta_m 
\end{array} \renewcommand{\arraystretch}{1}\right])\nonumber\\
&=&
\left[\begin{array}{@{}c@{\ }c@{\ }c@{}}
 \delta_1\colv{x}{1}  & \ldots & \delta_m\colv{x}{m} \end{array}\right]
 \label{WD}\\
 \diag (\left[ \renewcommand{\arraystretch}{.5}\begin{array}{@{}c@{}c@{}c@{}}
\delta_1, & \cdots, & \delta_m 
\end{array} \renewcommand{\arraystretch}{1}\right])\bd{\rm X} &=&\!\left[\begin{array}{@{}c@{\ }c@{\ }c@{}}
 \left( \delta_1\rowv{x}{1}\right)^{\rm T}   & {\ldots} & \left( \delta_m\rowv{x}{m}\right)^{\rm T}  \label{DW}
\end{array}\right]^{\rm T}
\end{eqnarray}
Apply \rf{DW}, \rf{WD} to $i$-th columns of $\bd{\rm H}^{\rm H} \bd{\rm H}\bd{\rm W} $ respectively $\bd{\rm W} \bd{\rm D}$ 
\be
\left[ 
\begin{array}{c}
\|\colv{h}{1}\|^2 w_{1i} \\
\vdots \\
\|\colv{h}{m}\|^2 w_{mi}
\end{array} 
\right] =
\left[ 
\begin{array}{c}
d_1 w_{1i} \\
\vdots \\
d_m w_{mi}
\end{array} 
\right]=
d_i\left[ 
\begin{array}{c}
 w_{1i} \\ 
\vdots \\
 w_{mi}
\end{array} 
\right]
\ee
If $\exists (\bd{\rm H}^{\rm H} \bd{\rm H})^{-1} $ and $d_i$ are 
all-distinct (eigenvalues), then  {\em necessarily} $ w_{ii}=1 $, and $ w_{ij}=0 $ $\forall i\neq j  $, i.e.\ $ \bd{\rm W} =\bd{\rm I}$ is the unique $m\times m$ matrix in the SVD. $ \bd{\rm W} =\bd{\rm I}$ remains a valid choice in the SVD, albeit not unique, even when $ \bd{\rm H}^{\rm H} \bd{\rm H} $ is singular or has  eigenvalues of multiplicity greater than one. Either way, 
$\bd{\rm H} = \bd{\rm V} \bd{\rm \Sigma}$
and $\exists$ a unitary matrix that diagonalizes $\bd{\rm H}$ on the left.
Conversely, if $\bd{\rm H} {=} \bd{\rm V} \bd{\rm \Sigma}$ then 
$ \bd{\rm H}^{\rm H} \bd{\rm H}{=}\bd{\rm \Sigma}^{\rm H}\bd{\rm V}^{\rm H} \bd{\rm V} \bd{\rm \Sigma} {=}  \bd{\rm \Sigma \Sigma}$. Q.E.D.
\hfill $\blacksquare$\\ 
The unconstrained LS solution to \rf{rxeq} is $\hat{\bd{s}}_{\rm ml}\!=\!\! \bd{\rm H}^{+}\!\bd{y}\!=\!\!(\bd{H}^{\rm H} \bd{H})^{-1} \bd{H}^{\rm H} \bd{\rm H} \bd{s}\!+\! \bd{\rm H}^{+}\!\bd{n}$, 
and $\exists$   $  \bd{\rm V}\!\! \in \!\mathbb{C}^{n\times n}$ unitary 
\ba
\bd{\chi}\df \bd{\rm V}^{\rm H}\bd{y}&=&\bd{\rm V}^{\rm H}\bd{\rm H} \bd{s}+\bd{\rm V}^{\rm H}\bd{n}=\gm{H}\bd{s}+\bd{n}^\prime \label{rxu}\\
\mathbb{C}^{n\times m} \ni \gm{H}&=&\left[ 
\begin{array}{@{}c@{}}
\diag (\gv{h}_{11}, \ldots, \gv{h}_{mm}) \\ 
\bd{0}_{(n-m)\times m}
\end{array} 
\right] 
\ea
and $\gv{h}_{ii}\geq 0$; \rf{rxeq} and \rf{rxu} are equivalent ($\bd{\chi}$ is a sufficient statistic), and  $\gm{H}$ has pseudoinverse $\gv{H}^{+}{=}(\gm{H}^{\rm H} \gm{H})^{-1} \gm{H}^{\rm H}$ 
\begin{eqnarray}
\gv{H}^{+}&\!\!\!{=}&
\!\!\!\diag\left(\left[ \begin{array}{@{}c@{}c@{}c@{}}
\gv{h}_{11}^{-2} & \ldots  & \gv{h}_{m m}^{-2} 
\end{array} \right] \right)
\left[ \begin{array}{@{}c@{}c@{}}
\diag\left(\left[ \begin{array}{@{}c@{}c@{}c@{}}
\gv{h}_{11} & \ldots  & \gv{h}_{m m} 
\end{array} \right]\right)
&
 \bd{0}_{m\times (n-m)}
\end{array} \right]\nonumber\\
&\!\!\!{=}&\!\!\!\left[ \begin{array}{@{}c@{}c@{}}
\diag\left(\left[ \begin{array}{@{}c@{}c@{}c@{}}
\gv{h}_{11}^{-1} & \ldots  & \gv{h}_{m m}^{-1} 
\end{array} \right] \right)&
 \bd{0}_{m\times (n-m)}
\end{array} \right]
\end{eqnarray}
Then the unconstrained LS solution is 
\be 
\hat{s}_{\rm ml}=\gm{H}^{+}\bd{\chi}=
\diag\left(\left[ \begin{array}{@{}c@{}c@{}c@{}}
\gv{h}_{11}^{-1} & \ldots  & \gv{h}_{m m}^{-1} 
\end{array} \right]\right)
\left[ \begin{array}{@{}c@{}c@{}c@{}}
\chi_1 & \cdots & \chi_m
\end{array}\right]^{\rm T} \label{unconstrML} 
\ee
The constrained ML solution $\hat{\bd{s}}_{\rm cML}{\doteq}\arg {\min}_{\bd{s} \in {\cal M}}\|\bd{\chi}{-}\bd{\rm H} \bd{s}\|^2$ 
\begin{eqnarray}
\hat{\bd{s}}_{\rm cML}&{=}&\textstyle \arg \underset{\bd{s} \in {\cal M}}{\min}\left\{ \sum_{i=1}^m |\chi_i -\gv{h}_{ii} s_i|^2 + \sum_{i=m+1}^n |\chi_i|^2\right\}\nonumber\\
&{=}&\textstyle \arg {\min}_{\bd{s} \in {\cal M}}\left\{ \sum_{i=1}^m |\gv{h}_{ii}|^2 |\gv{h}_{ii}^{-1} \chi_i-s_i|^2 \right\}\nonumber\\
&{=}&\textstyle \arg {\min}_{\bd{s} \in {\cal M}}\left\{ \sum_{i=1}^m |\gv{h}_{ii}|^2 |s_i -\hat{s}_{{\rm ml},i} |^2\right\} \label{optzf}
\end{eqnarray}
The minimum in \rf{optzf} corresponds to  the constrained  ML solution (closest $ \bd{s} {\in} {\cal M} $ to the sufficient statistic $\bd{\chi}$),  obtainable as the closest $ \bd{s} {\in} {\cal M} $ to the unconstrained LS solution $\hat{s}_{\rm ml}$ \rf{unconstrML}; 
should ${\cal M} $ be a lattice, rounding $\hat{s}_{\rm ml}$ 
to  nearest integers (Babai estimate) yields the exact integer LS solution to \rf{rxeq}. \hfill $\blacksquare$ 
}

\section*{Acknowledgment}
The authors gratefully acknowledge the professionalism and support of Dr.\ Ezio Biglieri, who generously  re-instated the manuscript after a   period of inactivity; of the new Associate Editor, Dr.\ Emanuele Viterbo---an established  expert in the topic of the manuscript---who kindly located and retrieved the old manuscript from the journal's archives, and was prompt,  helpful, yet patient enough in managing and expediting a new round of review; and of Dr.\ Helmut B\"olcskei for endorsing the re-instatement, in his role of new Editor-In-Chief. We also acknowledge the efforts and insightful suggestions of the referees. As a result of the altruistic 
peer review process it is the authors' opinion that the manuscript is greatly improved over its original version posted on the arXiv server in 2006.

\bibliographystyle{IEEE}

\begin{thebibliography}{99}

\bibitem{IonescuMYL:01} D.~M.\ Ionescu, K.~K.\ Mukkavilli, Z.~Yan, and J.~Lilleberg, ``{Improved 8- and 16-State Space-Time codes for 4PSK with Two Transmit Antennas},'' {\em {IEEE  Commun.\ Letters}}, vol.\ 5, pp.\ 301--303, July 2001.
\bibitem{ion:spa} D.\ M.\ Ionescu, ``{On space-time code design},'' {\em {IEEE  Trans.\ Wireless Commun.}}, vol.\ 2, pp.\ 20--28, Jan.\ 2003.
\bibitem{fer:new} G.\ Ferr\'{e} and J.-P.\ Cances, ``{New layered space-time schemes based on group of three transmit antennas},'' {\em {IEEE  Commun.\ Letters}}, vol.\ 12, pp.\ 38--40, Jan.\ 2008.
\bibitem{gha:div} M.\ Gharavi-Alkhansari and A.\ B.\ Gershman, ``On diversity and coding gains and optimal matrix constellations for space-time block codes,'' {\em {IEEE Trans.\ Signal Process.}},  vol.\ 53, pp.\ 3703--3717, Oct.\ 2005.
\bibitem{siw:impcon} S.\ Siwamogsatham and M.\ P.\ Fitz, ``Improved High-Rate Space-Time Codes via Concatenation of Expanded Orthogonal Block Code and M-TCM,'' \textit{Proc.\ 2002 ICC}, vol.\ 1, pp.\ 636-640, May 2002.
\bibitem{siw:import} S.\ Siwamogsatham and M.\ P.\ Fitz, ``Improved High-Rate Space-Time Codes via Orthogonality and Set Partitioning,'' \textit{Proc.\ 2002 IEEE WCNC}, vol.\ 1, pp.\ 264-270, March 2002.
\bibitem{ses:sup} N.\ Seshadri and H.\ Jafarkhani, ``Super-Orthogonal Space-Time Trellis Codes,'' in \textit{Proc.\ ICC'02}, May 2002, Vol.\ 3, pp.\ 1439-1443.
\bibitem{jaf:sup} H.\ Jafarkhani and N.\ Seshadri, ``Super-orthogonal space-time trellis codes,'' \textit{IEEE Trans.\ Inf.\ Theory}, vol.\ 49, pp.\ 937-950, Apr.\ 2003.
\bibitem{ion:fad} D.\ M.\ Ionescu and Z.\ Yan, ``Fading-resilient super-orthogonal space-time signal sets: Can good constellations survive in fading?,'' {\em IEEE Trans.\ Infom.\ Theory}, vol.\ 53, pp.\ 3219-3225, Sep.\ 2007.
\bibitem{tir:squ} O.\ Tirkkonen and A.\ Hottinen, ``Square-matrix embeddable space-time block codes for complex signal constellations,'' \textit{IEEE Trans.\ Inf.\ Theory}, vol.\ 48, pp.\ 384-395, Feb.\ 2002.
\bibitem{yan:geo} Z.\ Yan and D.\ M.\ Ionescu, ``Geometrical Uniformity of  a Class of Space-Time Trellis Codes,'' {\it IEEE Trans.\ Inf.\ Theory}, vol.\ 50, pp.\ 3343--3347, Dec.\ 2004.
\bibitem{gam:lat} H.\ El Gamal, G.\ Caire, M.\ O.\ Damen, ``Lattice coding and   decoding achieve optimal diversity-multiplexing tradeoff of MIMO channels,'' {\em IEEE Trans.\  Inf.\ Theory}, vol.\ 50, no.\ 6, pp.\ 968-985, June 2004.
\bibitem{has:hig} B.\ Hassibi and B.\ M.\ Hochwald, ``High-rate codes that are linear in space and time,'' {\em IEEE Trans.\ Inf.\ Theory,} vol.\ 48, pp.\ 1804-1824, July 2002.
\bibitem{far:ada} B.\ Farhang-Boroujeny,  \textit{Adaptive Filters: theory  and  applications.} Chichester, West Sussex, England: Wiley, 2000.
\bibitem{xiaodong:mmse}  X.\ Wang and H.\ V.\ Poor,   ``Iterative (turbo) soft interference cancellation and decoding forcoded CDMA'' {\it IEEE Trans. Commun.}, vol.\ 47, no.\ 7, pp.\ 1046-1061, Jul.\ 1999.
\bibitem{tuc:tur}  M.\ T\"chler, R.\ Koetter and A.\ C.\ Singer,   ``Turbo equalization: principles and new results,'' {\it IEEE Trans. Commun.}, vol.\ 50, no.\ 5, pp.\ 754-767, May 2002.
\bibitem{Low:Mat} M.\ C.\ Davey and D.\ MacKay, ``Low-Density Parity Check Codes over GF($q$),'' {\em IEEE Commun.\ Letters}, vol.\ 2, June 1998
\bibitem{BenedettoDMP:97} S.~Benedetto, D.~Divsalar, G.~Montorsi, and F.~Pollara, ``{A soft-input   soft-output APP module for iterative decoding of concatenated codes},'' {\em   {IEEE Commun.\ Lett.}}, vol.~1, pp.\ 22--24, Jan.\ 1997.
\bibitem{bah:opt} L.\ R.\ Bahl, J.\ Cocke, F.\ Jelinek, and J.\ Raviv, ``Optimal decoding of linear codes for minimizing symbol error rate,'' \textit{IEEE Trans.\ Inf.\ Theory}, vol.\ IT-20, pp.\ 284-287, Mar. 1974.
\bibitem{ion:int} D.\ M.\ Ionescu, D.\ Doan, S.\ Gray, ``On Interleaving Techniques for MIMO Channels and Limitations of Bit Interleaved Coded Modulation,'' [Online]. Available:  http://www.arxiv.org/abs/cs.IT/0510072.
\bibitem{dis:zhu} H.\ Zhu ``Multiple-input multiple-output data detection and channel estimation in flat fading environments," Ph.D.\ dissertation, Dept.\ Elect.\ Eng.,  Univ.\ of Utah, Salt Lake, UT, 2006.
\bibitem{rob:com} P.\ Robertson, E.\ Villebrun, and P.\ Hoeher, ``A comparison of optimal and suboptimal MAP decoding algorithms operating in the log domain,'' in \textit{Proc.\ Int.\ Conf.\ Communications}, June 1995, pp.\ 1009–1013.
\bibitem{dam:lat} M.\ O.\ Damen, A.\ Chkeif, and J.-C.\ Belfiore, ``Lattice code decoder for space–time codes,'' IEEE Commun.\ Lett., pp.\ 161–-163, May 2000.
\bibitem{Hochwald:LSD} B.\ M.\ Hochwald and S.\ ten\ Brink, ``Achieving near-capacity on a multiple antenna channel,'' \textit{IEEE Trans.\ Commun.}, vol.\ 51, pp.\ 389-399, Mar.\ 2003.
\bibitem{sad:low} M.\ R.\ Sadeghi, A.\ H.\ Banihashemi and D.\ Panario, ``Low density parity check lattices: construction and performance analysis,'' \textit{IEEE Trans.\ Inf.\ Theory}, pp.\ 4481-4495, Oct.\ 2006.
\bibitem{kan:imp} R.\ Kannan, ``Improved algorithms on integer programming and related lattice problems,'' in \textit{Proc.\ 15th Annu.\ ACM Symp.\ Theory Comput.}, 1983, pp.\ 193--206.
\bibitem{lag:kor} J.\ Lagarias, H.\ Lenstra, and C.\ Schnorr, ``Korkin-Zolotarev bases and successive minima of a lattice and its reciprocal,'' \textit{Combinatorica}, vol.\ 10, pp.\ 333--348, 1990.
\bibitem{poh:com} M.\ Pohst, ``On the computation of lattice vectors of minimal length, successive minima and reduced basis with applications,'' \textit{ACM SIGSAM Bull.}, vol.\ 15, pp.\ 37--44, 1981.
\bibitem{fin:imp}  U.\ Fincke and M.\ Pohst, ``Improved methods for calculating vectors of short length in a lattice, including a complexity analysis,'' \textit{Math.\ Comput.}, vol.\ 44, pp.\ 463--471, Apr.\ 1985.
\bibitem{agr:clo} E.\ Agrell, T.\ Eriksson,  A.\ Vardy, and K.\ Zeger, ``Closest point search in lattices,'' \textit{IEEE Trans.\ Inf.\ Theory}, vol.\ 48, No.\ 2, pp.\ 2201--2214, Aug.\ 2002.
\bibitem{vit:uni1} E.\ Viterbo and J.\ Boutros, ``A universal lattice code decoder for fading channels,'' \textit{IEEE Trans.\ Inf.\ Theory}, vol.\ 45, pp.\ 1639–-1642, July 1999.
\bibitem{ban:dis} A.\ H.\ Banihashemi, ``Trellis structure and decoding complexity of lattices,'' Ph.D.\ dissertation, ECE Dept., Univ.\ Waterloo, Waterloo, ON, Canada, 1997.
\bibitem{ban:tan} A.\ H.\ Banihashemi and F.\ R.\ Kschischang, ``Tanner graphs for group block codes and lattices: Construction and complexity,'' {\em IEEE Trans.\ Inf.\ Theory}, vol.\ 47, pp.\ 822--834, Feb.\ 2001.
\bibitem{has:sph} B.\ Hassibi and H.\ Vikalo, ``On the sphere decoding algorithm I. Expected complexity,'' \textit{IEEE Trans.\ Signal Process.}, vol.\ 53, pp.\ 2806--2818, Aug.\ 2005.
\bibitem{vit:uni2} E.\ Viterbo and E.\ Biglieri, ``A universal lattice decoder,'' in \textit{14 Colloq.\ GRETSI}, pp.\ 611–-614 Juan-les-Pins, France, Sept.\ 1993.
\bibitem{bou:goo} J.\ Boutros, E.\ Viterbo, C.\ Rastello, and J.\ C.\ Belfiore, ``Good lattice constellations for both Rayleigh fading and Gaussian channels,'' \textit{IEEE Trans.\ Inf.\ Theory}, vol.\ 42, pp.\ 502--518, Mar.\ 1996.
\bibitem{bou:sof} J.\ Boutros, N.\ Gresset, L.\ Brunel, and M.\ Fossorier, ``Soft-input soft-output lattice sphere decoder for linear channels,'' in \textit{Proc.\ Globecom 2003}, pp.\ 1583--1587.
\bibitem{stu:sof} C.\ Studer, A.\ Burg, and H.\ B\"olcskei, ``Soft-output sphere decoding: Algorithms and VLSI implementation,'' \textit{IEEE J.\ Sel.\ Areas Commun.}, vol.\ 26, pp.\ 290-300, Feb.\ 2008.
\bibitem{for:uni} G.\ D.\ Forney, ``Geometrically uniform codes,'' \textit{IEEE Trans.\ Inf.\ Theory}, vol.\ 37, No.\ 5, pp.\ 1241--1260, Sep.\ 1991.
\bibitem{tra:per}  N.\ H.\ Tran, H.\ H.\ Nguyen, and T.\ Le-Ngoc, ``{Performance of BICM-ID with signal space diversity},'' {\em {IEEE  Trans.\ Wireless Commun.}}, vol.\ 6, pp.\ 1732--1742, May 2007.
\bibitem{vit:ful} E.\ Viterbo, \textit{Full diversity rotations}. [Online]. Available: http://www1.tlc.polito.it/~viterbo/rotations/rotations.html.
\bibitem{ogg:alg} F.\ Oggier and E.\ Viterbo, ``Algebraic number theory and code design for Rayleigh fading channels,''  
 in \textit{Foundations and Trends in Communications and Information Theory}, vol.\ 1, pp.\ 333--415, 2004.
\bibitem{wan:sys} G.\ Wang, H.\ Liao, H.\ Wang, and X.-G.\ Xia, ``Systematic and optimal cyclotomic lattices and diagonal space-time block code designs,'' \textit{IEEE Trans.\ Inf.\ Theory}, vol.\ 50, pp.\ 3348--3360, Dec.\ 2004.
\bibitem{bou:sig} J.\ Boutros and E.\ Viterbo, ``Signal space diversity: A power and bandwidth efficient diversity technique for the Rayleigh fading channel,''
\textit{IEEE Trans.\ Inf.\ Theory}, vol.\ 44, pp.\ 1453--1467, July 1998.
\bibitem{tar:spa} V.\ Tarokh, N.\ Seshadri, and A.\ R.\ Calderbank, ``Space–time codes for high data rate wireless communication: Performance criteria and code
construction,'' \textit{IEEE Trans.\ Inf.\ Theory}, vol.\ 44, pp.\ 744--765, Mar.\ 1998.
\bibitem{bou:bit} J.\ J.\ Boutros, F.\ Boixadera, and C.\ Lamy, ``Bit-interleaved coded modulations for multiple-input multiple-output channels,'' \textit{Proc.\  IEEE 6th Int.\ Symp.\ on Spread-Spectrum Tech.\ \& Applicat.,}, pp.\ 123--126,  Sep.\ 6--8,  2000.
\bibitem{ion:app} D.\ M.\ Ionescu and A.\ Reid, ``Apparatus using concatenations of signal-space codes for jointly encoding across multiple transmit antennas, and employing coordinate interleaving,'' U.S.\ Patent ---, Oct.\ 4, 2011. [Online]. Available: http://patft.uspto.gov/ (pre-grant publication 20060159195).
\bibitem{gro:geo} M.\ Grotschel, L.\ Lovász, and A.\ Schriver, \textit{Geometric Algorithms and Combinatorial Optimization}, 2nd ed.\ New York: Springer-Verlag, 1993.
\bibitem{has:eff} B.\ Hassibi, ``An efficient square-root algorithm for BLAST,'' in \textit{2000 IEEE Int.\ Conf.\ Acoust.,  Speech, Signal Process.\ Proc.}, vol.\ 2, pp.\ II737--II740.
\bibitem{fos:lay} G.\ J.\ Foschini, ``Layered space-time architecture for wireless communication in a fading environment when using multi-element antennas,''
Bell Labs.\ Tech.\ J., vol.\ 1, no.\ 2, pp.\ 41--59, 1996.
\bibitem{mon:des} G.\ Montorsi and S.\ Benedetto, ``Design of fixed-point iterative decoders for concatenated codes with interleavers,'' \textit{IEEE J.\ Sel.\ Areas Commun.}, vol.\ 19, pp.\ 871--881, May 2001.
\bibitem{mud:min} D.\ J.\ Muder, ``Minimal trellises for block codes,'' \textit{IEEE Trans.\ Inf.\ Theory}, vol.\ 34, No.\ 5, pp.\ 1049-1053, Sep.\ 1988.
\bibitem{laf:low} A.\ Lafourcade and A.\ Vardy, ``Lower bounds on trellis complexity of block codes,'' \textit{IEEE Trans.\ Inf.\ Theory}, vol.\ 41, No.\ 6, pp.\ 1938--1954, Nov.\ 1995.
\bibitem{ytr:tre} \O.\ Ytrehus, ``On the trellis complexity of certain binary linear block codes,'' \textit{IEEE Trans.\ Inf.\ Theory}, vol.\ 41, No.\ 2, pp.\ 559-560, Mar.\ 1995.
\bibitem{laf:asy} A.\ Lafourcade and A.\ Vardy, ``Asymptotically good codes have infinite trellis complexity,'' \textit{IEEE Trans.\ Inf.\ Theory}, vol.\ 41, No.\ 2, pp.\ 555--559, Mar.\ 1995.
\bibitem{mce:tre} R.\ J.\ McEliece, ``The trellis complexity of convolutional codes,'' Caltech, Pasadena, CA, TDA Progress Rep.\ 42-123, pp.\ 122--139, Nov.\ 15, 1995.
\bibitem{con:sph} J.\ H.\ Conway and N.\ J.\ A.\ Sloane, \textit{Sphere packings, lattices and groups}, 3rd ed. New York, NY: Springer-Verlag, 1999.
\bibitem{for:den} G.\ D.\ Forney Jr., ``Density/length profiles and trellis complexity of lattices,'' \textit{IEEE Trans.\ Inf.\ Theory}, vol.\ 40, pp.\ 1753--1772, Nov.\ 1994.
\bibitem{ban:tre} A.\ H.\ Banihashemi and I.\ F.\ Blake, ``On the trellis complexity of root lattices and their duals,'' {\em IEEE Trans.\ Inf.\ Theory}, vol.\ 45, No.\ 6, pp.\ 2168--2172, Sep.\ 1999.
\end{thebibliography}

\begin{biography}{Dumitru Mihai Ionescu}
(S'93-M'96-SM'04) received the M.Sc.\ degree in electrical
engineering from the Technical University ``Gheorghe Asachi," Iasi,
Romania, in 1986, and the Ph.D.\ degree in electrical engineering
from the University of Colorado, Colorado Springs, in 1996.

After teaching positions in the Electronics \& Telecommunications Department, Iasi Technical
University, from 1990 to 1992, and University of Colorado, 
he worked from 1996 to 1998 for
Omnipoint Corp., Colorado Springs, CO, where he was involved
in communications and signal processing development aspects of a
proprietary technology, IS-661. From 1998 to 2006 he held positions with Nokia Research Center, in
Irving, TX, then San Diego, CA, where he conducted research in the areas of
 communications theory,
modulation and coding, and adaptive algorithms. During his tenure with Olympus Communication Technology of America, Inc., in San Diego, CA, he was involved with developing and implementing ultra-wide band multi-band OFDM, and short-range and cognitive radios. During 2001, he
served as an Adjunct Assistant Professor in the Electrical
Engineering Department, Southern Methodist University, Dallas, TX.
His current research interests include modulation and coding,
turbo and space–time codes, lattice decoding, codes on graphs, (sparse) signal processing, and synchronization algorithms for multicarrier systems.

Dr.\ Ionescu served as Program Chair for the IEEE
Telecommunications Chapter, Fort Worth, TX, and as an AE for IEEE; he has been granted 
eighteen patents, including two for the so-called super-orthogonal space-time codes.
\end{biography}

\begin{biography}{Haidong (David) Zhu}
(S'03-M'06) received the B.S.\ and M.S.\
degrees in electrical engineering from Fudan University,
Shanghai, China, in 1998 and 2001, respectively, and the Ph.D.\ degree from
the University of Utah, Salt Lake City, in 2006.
 
He was a Research Assistant with
the University of Utah from 2001 to 2006. In summer 2005, he was a
Research Intern with Nokia Research Center, San 
Diego, CA. From 2006 to 2008, he was Sr.\ CDMA system engineer with Motorola mobile device in San Diego. From 2008 to 2011, 
 he was member of technical staff with Olympus communication technology of America in San Diego. Currently, he is with Qualcomm QCT RFA group in San Diego, CA. \end{biography}
\end{document}